\documentclass[aps,prb,twocolumn,unsortedaddress,showpacs]{revtex4-1}
\usepackage{amsmath}
\usepackage{bm}
\usepackage{graphicx}
\usepackage{caption}
\usepackage{subcaption}
\usepackage{feynmp}
\usepackage{multirow}
\usepackage{hyperref}
\usepackage{jabbrv}

\newcommand{\dd}{\mathrm{d}}
\newcommand{\6}{\partial}
\newcommand{\mbf}{\mathbf}

\newcommand{\mrm}{\mathrm}

\newcommand{\eps}{\varepsilon}
\usepackage{color}

\begin{document}

\title{Hubbard-like Hamiltonians for interacting electrons in s, p and d orbitals}
\author{M.~E.~A.~Coury}
\email{marc.coury07@imperial.ac.uk}
\affiliation{Imperial College London, London SW7 2AZ, United Kingdom}
\affiliation{CCFE, Culham Science Centre, Abingdon, Oxfordshire OX14 3DB, United Kingdom}
\author{S.~L.~Dudarev}
\affiliation{CCFE, Culham Science Centre, Abingdon, Oxfordshire OX14 3DB, United Kingdom}
\author{W.~M.~C.~Foulkes}
\affiliation{Imperial College London, London SW7 2AZ, United Kingdom}
\author{A.~P.~Horsfield}
\affiliation{Imperial College London, London SW7 2AZ, United Kingdom}
\author{Pui-Wai Ma}
\affiliation{CCFE, Culham Science Centre, Abingdon, Oxfordshire OX14 3DB, United Kingdom}
\author{J.~S.~Spencer}
\affiliation{Imperial College London, London SW7 2AZ, United Kingdom}

\begin{abstract}
Hubbard-like Hamiltonians are widely used to describe on-site Coulomb interactions in magnetic and strongly-correlated solids, but there is much confusion in the literature about the form these Hamiltonians should take for shells of p and d orbitals. This paper derives the most general s, p and d orbital Hubbard-like Hamiltonians consistent with the relevant symmetries, and presents them in ways convenient for practical calculations. We use the full configuration interaction method to study p and d orbital dimers and compare results obtained using the correct Hamiltonian and the collinear and vector Stoner Hamiltonians. The Stoner Hamiltonians can fail to describe properly the nature of the ground state, the time evolution of excited states, and the electronic heat capacity.
\end{abstract}
\maketitle

\section{Introduction}
\label{s:intro}
The starting point for any electronic structure calculation is the Hamiltonian that describes the dynamics of the electrons. Correlated electron problems are often formulated using Hamiltonians that involve only a small number of localized atomic-like orbitals and use an effective Coulomb interaction between electrons that is assumed to be short ranged, and thus confined to within atomic sites. However, there is widespread variation in the literature over the form chosen for these Hamiltonians, and they often do not retain the correct symmetry properties. Here we derive the most general Hamiltonians of this type that are rotationally invariant and describe electrons populating s, p and d orbitals.

Despite often missing terms, multi-orbital Hamiltonians with on-site Coulomb interactions have been used successfully to describe strong correlation effects in systems with narrow bands near the Fermi energy. Applications include studies of metal-insulator transitions \cite{imada1998}, colossal magnetoresistant manganites \cite{dagotto2001}, and d band superconductors such as copper oxides \cite{zhang1988} and iron pnictides \cite{zhou2011}. The Hubbard Hamiltonian has also been used to simulate graphene \cite{wakabayashi1998,wehling2011}, where both theory and experiment have found magnetic ordering in nanoscale structures \cite{shibayama2000,enoki2009}.  However, the lack of consistency in the forms used for the Hamiltonians is problematic. Indeed, according to Dagotto \cite{dagotto2001}, ``the discussions in the current literature regarding this issue are somewhat confusing''.

The Hubbard Hamiltonian \cite{hubbard1963} is both the simplest and best known Hamiltonian of the form we are considering, and is valid for electrons in s orbitals. Hubbard was not the first to arrive at this form; two years earlier, in 1961, Anderson wrote down a very similar Hamiltonian \cite{anderson1961} for electrons in s orbitals with on-site Coulomb interactions. Furthermore, in the appendix of the same paper, the Hamiltonian was extended to two orbitals, introducing the on-site exchange interaction, $J$. Unfortunately, this extra term is not rotationally invariant in spin space.

In 1966, Roth \cite{roth1966} also considered two orbitals and two parameters, producing a Hamiltonian that is rotationally invariant in spin space. In 1969 Caroli \emph{et al.}~\cite{caroli1969} corrected Anderson's Hamiltonian for two orbitals, again making it rotationally invariant in spin space. However, the Hamiltonians of Roth and Caroli \emph{et al.}~do not satisfy rotational invariance in orbital space. This was corrected by Dworin and Narath in 1970 \cite{dworin1970}, who produced a Hamiltonian that is rotationally invariant in both spin and orbital space. They generalized the multi-orbital Hamiltonian to include all 5 d orbitals, although they only included two parameters: the on-site Hartree integral, $U$, and the on-site exchange integral, $J$.

The next contribution to the multi-orbital Hamiltonian was by Lyon-Caen and Cyrot \cite{lyon1975}, who in 1975 considered a two-orbital Hamiltonian for the $e_g$ d orbitals. They introduced $U_0 -U = 2J$, where $U_0=V_{\alpha\alpha,\alpha\alpha}$ is the self-interaction term, $U=V_{\alpha\beta,\alpha\beta}$ is the on-site Hartree integral, and $J = V_{\alpha\beta,\beta\alpha}$ is the on-site exchange integral, important for linking these parameters. Here $V$ is the Coulomb interaction between electrons, and $\alpha$ and $\beta$ are indices for atomic orbitals. Lyon-Caen and Cyrot also included the pair excitation, $J\hat{c}^{\dagger}_{I1,\uparrow}\hat{c}^{\dagger}_{I1,\downarrow}\hat{c}_{I2,\downarrow}\hat{c}_{I2,\uparrow}$, where $\hat{c}^{\dagger}$ and $\hat{c}$ are fermionic creation and annihilation operators. The new term moves a pair of electrons from a completely occupied orbital, 2, on site $I$, to a completely unoccupied orbital, 1, on site $I$. Such double excitations are common in the literature \cite{kanamori1963,moszynski1994,fast2000,maitra2004}. 

In 1978, Castellani \emph{et al.}~\cite{castellani1978} wrote down the three-orbital Hamiltonian for the $t_{2g}$ d orbitals\footnote{In this notation $t_{2g}$ refers to cubic harmonic d orbitals $d_{xy}$, $d_{yz}$ and $d_{xz}$ and $e_g$ refers to cubic harmonic d orbitals $d_{3z^2-r^2}$ and $d_{x^2-y^2}$.} in a very clear and concise way: its form is exactly what we find here for p orbital symmetry. They too make note of the equation $U_0 -U = 2J$ and include the double excitation terms.

The next major contribution was by Ole\'{s} and Stollhoff \cite{oles1984}, who introduced a d orbital Hamiltonian with three independent parameters. The third parameter, $\Delta J$, represents the difference between the $t_{2g}$ and the $e_g$ d orbital exchange interactions; they estimated $\Delta J$ to be $0.15 J$. In our discussion of d orbitals we adopt their notation. Unfortunately, their Hamiltonian is not rotationally invariant in orbital space.

In 1989, Nolting \emph{et al.}~\cite{nolting1989} referenced the Hamiltonian proposed by Ole\'{s} and Stollhoff but discarded the double excitation terms and ignored the $\Delta J$ term. Their Hamiltonian will be called the vector Stoner Hamiltonian in this paper. Since then, most papers have used a Hamiltonian similar to that of Nolting \emph{et al.},~with only two parameters; see for example references \onlinecite{dudarev2007,barreteau2007,duc2009}. The vector Stoner Hamiltonian is often simplified by replacing the rotationally invariant moment-squared operator, $\hat{\mbf{m}}^2 = \hat{\mbf{m}}\cdot\hat{\mbf{m}}$, by $\hat{m}_z^2$, yielding the collinear Stoner Hamiltonian.

The multi-orbital Hamiltonians derived here consist of one-particle hopping (inter-site) integrals and two-electron on-site Coulomb interactions. The on-site Coulomb interactions for s, p and d orbital symmetries are presented in a clear tensor notation, as well as in terms of physically meaningful, rotationally invariant (in both orbital and spin space), operators that include all of the terms from the previous papers as well as additional terms of the same order of magnitude that have not been previously considered. The d orbital Hamiltonian presented in section \ref{ss:dcase} corrects the d orbital Hamiltonian proposed by Ole\`{s} and Stollhoff \cite{oles1984} by restoring full rotational invariance in orbital space. For an s orbital Hamiltonian, the on-site Coulomb interactions may be described using one independent parameter. Two independent parameters are required for p orbital symmetry, and three for d orbital symmetry. Furthermore, the method presented can be extended to f and g orbital symmetry and generalized to atoms with valence orbitals of multiple different angular momenta.

The use of a restricted basis set and the neglect of the two-electron inter-site Coulomb interactions means that screening effects are missing. Thus, even though the parameters that define the s, p and d Hamiltonians appear as bare on-site Coulomb integrals, screened Coulomb integrals should be used when modelling real systems; this significantly reduces the values of the parameters.

Although it is possible to find the on-site Coulomb integrals using tables of Slater-Condon parameters\cite{slater1929,condon1930} or the closed forms for the integrals of products of three spherical harmonics found by Gaunt \cite{gaunt1929} and Racah \cite{racah1942}, the point of this paper is to remove that layer of obscurity and present model Hamiltonians written in a succinct form with clear physical meaning. The Slater-Condon parameters and Gaunt integrals were not used in the derivations of the on-site Coulomb interaction Hamiltonians presented here; however, they did provide an independent check to confirm that the tensorial forms derived here are correct. The link between the spherical harmonics Coulomb integrals, the Slater-Condon parameters and the Racah parameters is explained clearly in \cite{powell1998physics}. The transformation to cubic harmonics is straightforward; an example of which has been included for the p-orbitals in Section \ref{ss:pcase}.

We note that Rudzikas \cite{rudzikas1997} has derived rotationally invariant Hamiltonians for p and d orbital atoms. However, his derivation is different from ours and his d orbital Hamiltonian is partially expressed in terms of tensors; the form we present in this paper is written in terms of physically meaningful operators.

The p and d symmetry Hamiltonians are described in Section \ref{s:theory} (see Appendix \ref{append:representation} for full derivations) and compared with the vector Stoner Hamiltonian. In Section \ref{s:GS} we show the ground state results for both our Hamiltonian and the vector Stoner Hamiltonian for p and d orbital dimers. We find that, although the results are often similar, there are important qualitative and quantitative differences. In Section \ref{s:excitations} we show the differences that emerge when evolving a starting state for a p orbital dimer under our Hamiltonian and the vector Stoner Hamiltonian. We also calculate a thermodynamic property, the electronic heat capacity of a p orbital dimer, and obtain significantly different results when using our Hamiltonian and the vector Stoner Hamiltonian. The eigenstates of all Hamiltonians are found using the full configuration interaction (FCI) method as implemented in the HANDE code \cite{HANDE}.

\section{Theory}
\label{s:theory}
\subsection{General form of the Hamiltonian}
\label{ss:theory_general}
We start with the general many-body Hamiltonian of electrons in a solid, expressed in second-quantized form using a basis of localized one-electron orbitals and ignoring terms that do not include electronic degrees of freedom\cite{helgaker2001}:
\begin{align}
    \nonumber   
        \hat H =& \sum_{I\alpha J\beta}\sum_{\sigma}t^{\phantom{\dagger}}_{I\alpha J\beta}\hat{c}^{\dagger}_{I\alpha,\sigma}\hat{c}^{\phantom{\dagger}}_{J\beta,\sigma} \\ 
    \label{eq:2ndquantisationHamilocal}
        &+\sum_{IJKL}\sum_{\alpha\beta\chi\gamma}\sum_{\sigma\xi}V^{\phantom{\dagger}}_{I\alpha J\beta, K\chi L\gamma} \hat{c}^{\dagger}_{I\alpha,\sigma}\hat{c}^{\dagger}_{J\beta,\xi}\hat{c}^{\phantom{\dagger}}_{L\gamma,\xi}\hat{c}^{\phantom{\dagger}}_{K\chi,\sigma}.
\end{align}
Here $\hat{c}^\dagger_{I\alpha,\sigma}$ and $\hat{c}^{\phantom{\dagger}}_{I\alpha,\sigma}$ are creation and annihilation operators, respectively, for an electron in orbital $\alpha$ on site $I$ with spin $\sigma$. Upper case Roman letters such as $I$, $J$, $K$ and $L$ refer to atomic sites, the lower case Greek letters $\alpha$, $\beta$, $\chi$ and $\gamma$ represent localized orbitals, and $\sigma$ and $\zeta$ indicate spin (either up or down). The one-electron contributions to the Hamiltonian are encapsulated in the $t_{I\alpha J\beta}$ term that includes the electronic kinetic energy and electron-nuclear interaction. The electron-electron interaction terms are represented by the Coulomb matrix elements $V_{I\alpha J\beta, K\chi L\gamma}$.

We now make two approximations: we only retain electron-electron Coulomb interactions on-site and we restrict the basis to a minimal set of localized angular momentum orbitals. Thus we have just one s orbital per site for the s-symmetry Hamiltonian, three p orbitals per site for the p-symmetry Hamiltonian, and five d orbitals per site for the d-symmetry Hamiltonian. The Hamiltonian can then be written as follows:
\begin{align}
    \nonumber   
        \hat H \approx& \sum_{I\alpha J\beta}\sum_{\sigma}t^{\phantom{\dagger}}_{I\alpha J\beta}\hat{c}^{\dagger}_{I\alpha,\sigma}\hat{c}^{\phantom{\dagger}}_{J\beta,\sigma} \\
    \label{eq:2ndquantisationHamilocal_onsite}
        &+ \frac{1}{2}\sum_{I}\sum_{\alpha\beta\chi\gamma}\sum_{\sigma\xi}V^I_{\alpha \beta, \chi\gamma} \hat{c}^{\dagger}_{I\alpha,\sigma}\hat{c}^{\dagger}_{I\beta,\xi}\hat{c}^{\phantom{\dagger}}_{I\gamma,\xi}\hat{c}^{\phantom{\dagger}}_{I\chi,\sigma},
\end{align}
where $V^I_{\alpha \beta, \chi \gamma}=V^{\phantom{I}}_{I\alpha I\beta, I\chi I\gamma}$. To improve readability, from this point onwards we drop the site index, $I$, from the Coulomb integrals and creation and annihilation operators as we shall always be referring to on-site Coulomb interactions. 

The on-site Coulomb interaction, $V_{\alpha\beta,\chi\gamma}$, is a rotationally invariant tensor if the atomic-like orbitals are angular momentum eigenstates. To demonstrate this we note that applying the rotation operator $\hat{\mathbf{R}}$ to an atomic-like orbital $\phi_{\alpha}$ of angular momentum $\ell$ gives
\begin{align}
        \hat{\mathbf{R}}|\phi^{\phantom{\ell}}_{\alpha}\rangle  
        & = \sum_{\beta}|\phi^{\phantom{\ell}}_{\beta}\rangle d^{\ell}_{\beta\alpha}(\mbf{R}),
    \label{eq:rotation}
\end{align}
where $d^{\ell}_{\beta\alpha}({\mbf{R}})$ is the $(2\ell+1)\times(2\ell+1)$ matrix that corresponds to the rotation $\hat{\mbf{R}}$ in the irreducible representation of angular momentum $\ell$. By making the change of variable $\mbf{x} = \hat{\mathbf{R}}\mbf{r}$ and $\mbf{x}' = \hat{\mathbf{R}}\mbf{r}'$, it is straightforward to show that
\begin{align}
    \nonumber
        &V_{\alpha\beta,\chi\gamma} = \int \dd\mathbf{r}\dd\mathbf{r}'\frac{\phi^*_{\alpha\sigma}(\mathbf{r})\phi^*_{\beta\sigma'}(\mathbf{r}')\phi^{\phantom{*}}_{\chi\sigma}(\mathbf{r})\phi^{\phantom{*}}_{\gamma\sigma'}(\mathbf{r}')}{|\mathbf{r}-\mathbf{r}'|},\\
    \label{eq:V_abcd_rotinvar}
        & = \left ( d^{\ell}_{\alpha'\alpha}(\mbf{R}) \right )^{*} \left ( d^{\ell}_{\beta'\beta}(\mbf{R}) \right )^{*} V^{\phantom{\ell}}_{\alpha'\beta',\chi'\gamma'}d^{\ell}_{\chi'\chi}(\mbf{R})d^{\ell}_{\gamma'\gamma}(\mbf{R}),
\end{align}
which demonstrates that $V_{\alpha\beta,\chi\gamma}$ is indeed a rotationally invariant tensor.

\subsection{The one band Hubbard model: s orbital symmetry}
\label{ss:scase}
The Hubbard Hamiltonian is applicable only to one band models where the one orbital per atom has s symmetry. This gives a simple form for the on-site Coulomb interaction tensor, $U_0 = V_{\alpha \alpha, \alpha \alpha}$; as there is only one type of orbital, there is only one matrix element. We can use this result to simplify the Coulomb interaction part of Eq.~(\ref{eq:2ndquantisationHamilocal_onsite}) and express it in terms of the electron number operator or the magnetic moment vector operator, defined as follows:
\begin{align}
    \label{eq:def_num_op}
        \hat{n} & = \sum_{\alpha\zeta} \hat{c}^{\dagger}_{\alpha,\zeta}\hat{c}^{\phantom{\dagger}}_{\alpha,\zeta},\\
    \label{eq:def_mag_op}
        \hat{\mbf{m}} & = \sum_{\alpha\zeta\zeta'} \hat{c}^{\dagger}_{\alpha,\zeta}\bm{\sigma}^{\phantom{\dagger}}_{\zeta\zeta'}\hat{c}^{\phantom{\dagger}}_{\alpha,\zeta'},\\
    \nonumber
        \bm{\sigma}^{\phantom{x}}_{\zeta\zeta'} & = (\sigma^{x}_{\zeta\zeta'},\sigma^{y}_{\zeta\zeta'},\sigma^{z}_{\zeta\zeta'}),
\end{align}
where $\sigma$ are the Pauli spin matrices and the sum over $\alpha$ has only one term as there is only one spatial orbital for the s case.
Hence
\begin{equation}
    \label{eq:C_op_num_scase} 
        \frac{1}{2}\sum_{\alpha\beta\chi\gamma}\sum_{\sigma\xi}V^{\phantom{\dagger}}_{\alpha \beta, \chi\gamma} \hat{c}^{\dagger}_{\alpha,\sigma}\hat{c}^{\dagger}_{\beta,\xi}\hat{c}^{\phantom{\dagger}}_{\gamma,\xi}\hat{c}^{\phantom{\dagger}}_{\chi,\sigma} = \frac{1}{2} U_0 :\hat{n}^2:,\\
\end{equation}
or, equivalently,
\begin{equation}   
    \label{eq:C_op_mag_scase}
        \frac{1}{2}\sum_{\alpha\beta\chi\gamma}\sum_{\sigma\xi}V^{\phantom{\dagger}}_{\alpha \beta, \chi\gamma} \hat{c}^{\dagger}_{\alpha,\sigma}\hat{c}^{\dagger}_{\beta,\xi}\hat{c}^{\phantom{\dagger}}_{\gamma,\xi}\hat{c}^{\phantom{\dagger}}_{\chi,\sigma} = -\frac{1}{6} U_0 :\hat{\mbf{m}}^2:,
\end{equation}
where we have use the normal ordering operator, $::$, to remove self interactions. The action of this operator is to rearrange the creation and annihilation operators such that all the creation operators are on the left, without adding the anticommutator terms that would be required to leave the product of operators unaltered; if the rearrangement requires an odd number of flips, the normal ordering also introduces a sign change. For example
\begin{align}
    \nonumber
        :\hat{n}^2: &= \sum_{\alpha\beta\sigma\zeta}:\hat{c}^{\dagger}_{\alpha,\zeta}\hat{c}^{\phantom{\dagger}}_{\alpha,\zeta}\hat{c}^{\dagger}_{\beta,\sigma}\hat{c}^{\phantom{\dagger}}_{\beta,\sigma}:,\\
    \label{eq:def_num_op_sq}
        & = \sum_{\alpha\beta\sigma\zeta} \hat{c}^{\dagger}_{\alpha,\zeta}\hat{c}^{\dagger}_{\beta,\sigma}\hat{c}^{\phantom{\dagger}}_{\beta,\sigma}\hat{c}^{\phantom{\dagger}}_{\alpha,\zeta}.
\end{align}
If the normal ordering operator is not used in Eqs.~(\ref{eq:C_op_num_scase}) and (\ref{eq:C_op_mag_scase}) then additional one electron terms have to be subtracted to remove the self interaction.

The mean-field form of this Hamiltonian has been included in Appendix \ref{append:scase_meanfield}.

\subsection{The multi-orbital model Hamiltonian: p orbital symmetry}
\label{ss:pcase}
Consider the case where the local orbitals, $\alpha$, $\beta$, $\chi$ and $\gamma$ appearing in Eq.~(\ref{eq:V_abcd_rotinvar}) are real cubic harmonic p orbitals with angular dependence $x/r$, $y/r$, and $z/r$. In this case the rotation matrices, $d^{\ell}_{\alpha'\alpha}(\mbf{R})$, are simply $3\times 3$ Cartesian rotation matrices. This makes $V_{\alpha\beta,\chi\gamma}$ a rotationally invariant fourth-rank Cartesian tensor, the general form of which is\cite{matthews2000},
\begin{equation}
    \label{eq:V_pcase_prime}
        V_{\alpha\beta,\chi\gamma} = U\delta_{\alpha\chi}\delta_{\beta\gamma}+J\delta_{\alpha\gamma}\delta_{\beta\chi}+J'\delta_{\alpha\beta}\delta_{\chi\gamma},
\end{equation}
where $ U = V_{\alpha\beta,\alpha\beta}$, $J = V_{\alpha\beta,\beta\alpha}$ and $J' = V_{\alpha\alpha,\beta\beta}$, with $\alpha\neq\beta$. By examination of the integral in Eq.~(\ref{eq:V_abcd_rotinvar}), we note that the Coulomb tensor has an additional symmetry when the orbitals are real (as are the cubic harmonic p orbitals), namely $V_{\alpha\beta,\chi\gamma}=V_{\chi\beta,\alpha\gamma}$ and $V_{\alpha\beta,\chi\gamma} = V_{\alpha\gamma,\chi\beta}$, and hence $J$ must be equal to $J'$. Thus we find
\begin{equation}
    \label{eq:V_pcase}
        V_{\alpha\beta,\chi\gamma} = U\delta_{\alpha\chi}\delta_{\beta\gamma}+J\left(\delta_{\alpha\gamma}\delta_{\beta\chi}+\delta_{\alpha\beta}\delta_{\chi\gamma}\right).
\end{equation}
This recovers the well known equation $U_0=U+2J$, where $U_0 = V_{\alpha\alpha,\alpha\alpha}$\cite{lyon1975, castellani1978}, and shows that the most general cubic p orbital interaction Hamiltonian is defined by exactly two independent parameters\cite{griffith1961}. In Appendix \ref{subappend:pcase} we give a full derivation of Eq.~(\ref{eq:V_pcase}) using representation theory.

In passing we observe that symmetric fourth-rank isotropic tensors can be found in other areas of physics, such as the stiffness tensor in isotropic elasticity theory\cite{kosevich2005}
\begin{equation}
    \label{eq:elasticity_tensor}
        C_{iklm} = \lambda\delta_{ik}\delta_{lm}+\mu(\delta_{il}\delta_{km}+\delta_{im}\delta_{kl}),
\end{equation}
where $\lambda$ is the Lam\'e coefficient and $\mu$ is the shear modulus.

Transforming Eq.~(\ref{eq:V_pcase}) into a basis of complex spherical harmonic p orbitals with angular dependence of the form $Y_{\ell m}(\theta,\phi)$ is straightforward. We write $V_{m m'',m' m'''}=\sum_{\alpha\beta\chi\gamma}l^*_{m\alpha}l^*_{m''\beta}l_{m'\chi}l_{m'''\gamma}V_{\alpha\beta,\chi\gamma}$, where $l_{m\alpha}$ is defined by
\begin{equation}
    \label{eq:c2s4p}
        p^{\mrm{sph}}_{m}=\sum_{\alpha}l_{m\alpha}p_\alpha,
\end{equation}
and has the following values
\begin{equation}
    \label{eq:c2s4pl}
        \left[l_{m\alpha}\right] =
 \left(
\begin{array}{ccc}
 \frac{1}{\sqrt{2}} & -\frac{i}{\sqrt{2}} & 0 \\
 0 & 0 & 1 \\
 -\frac{1}{\sqrt{2}} & -\frac{i}{\sqrt{2}} & 0 \\
\end{array}
\right).
\end{equation}
The result of applying this transformation is
\begin{align}
    \nonumber
        V_{m m'',m'm'''} =& U \delta_{m m'}\delta_{m'' m'''} + J \delta_{m m'''}\delta_{m'' m'}\\
    \label{eq:V_sphporb_tensor}
        &+(-1)^{m+m'}J\delta_{-m m''}\delta_{-m' m'''},
\end{align}
where $m$, $m'$, $m''$ and $m'''$ are spherical harmonics indices going from $-1$ to $+1$, $U = V_{10,10}$, and $J=V_{10,01}$. 

Using Eq.~(\ref{eq:V_pcase}) as a convenient starting point, it is straightforward to rewrite the on-site Coulomb Hamiltonian in terms of rotationally invariant operators:\cite{rudzikas1997}
\begin{align}
    \nonumber
        \frac{1}{2}\sum_{\alpha\beta\chi\gamma}\sum_{\sigma\xi}&V^{\phantom{\dagger}}_{\alpha \beta, \chi\gamma} \hat{c}^{\dagger}_{\alpha,\sigma}\hat{c}^{\dagger}_{\beta,\xi}\hat{c}^{\phantom{\dagger}}_{\gamma,\xi}\hat{c}^{\phantom{\dagger}}_{\chi,\sigma} \\
    \label{eq:C_pcase_1}
        = \frac{1}{2} & \Bigg(\left(U-J\right) :\hat{n}^2:-J:\hat{\mathbf{m}}^2:-J:\mathbf{L}^2:\Bigg),
\end{align}
where the vector angular momentum operator for p orbitals on site $I$ is defined to be
\begin{equation}
    \label{eq:angular_momentum_pcase}
        \hat{\mathbf{L}} = i\sum_{\alpha\beta\sigma}(\epsilon^{\phantom{\dagger}}_{1 \beta\alpha},\epsilon^{\phantom{\dagger}}_{2 \beta\alpha},\epsilon^{\phantom{\dagger}}_{3 \beta\alpha})\hat{c}^\dagger_{\alpha,\sigma}\hat{c}^{\phantom{\dagger}}_{\beta,\sigma},
\end{equation}
and $\epsilon_{\mu\beta\alpha}$ is the three-dimensional Levi-Civita symbol. An equivalent expression is
\begin{align}
    \nonumber
        \frac{1}{2}\sum_{\alpha\beta\chi\gamma}\sum_{\sigma\zeta}&V^{\phantom{\dagger}}_{\alpha \beta, \chi\gamma} \hat{c}^{\dagger}_{\alpha,\sigma}\hat{c}^{\dagger}_{\beta,\zeta}\hat{c}^{\phantom{\dagger}}_{\gamma,\zeta}\hat{c}^{\phantom{\dagger}}_{\chi,\sigma} \\
    \nonumber
        = \frac{1}{2} & \Bigg(\left(U-\frac{1}{2}J\right) :\hat{n}^2:-\frac{1}{2}J:\hat{\mathbf{m}}^2:\\
    \label{eq:C_pcase_2}
        &+J\sum_{\alpha\beta}:\left(\hat{n}_{\alpha\beta}\right)^2:\Bigg),
\end{align}
where the final term corresponds to on-site electron hopping\cite{sakai2006},
\begin{equation}
    \label{eq:def_n_ab}
        \hat{n}_{\alpha\beta} = \sum_{\sigma}\hat{c}^\dagger_{\alpha,\sigma}\hat{c}^{\phantom{\dagger}}_{\beta,\sigma}.
\end{equation}
Eq.~(\ref{eq:C_pcase_1}) embodies Hund's rules for an atom. By making the substitution $\hat{\mathbf{m}}=2\hat{\mathbf{S}}$, we see that the spin is maximized first (prefactor $-2J$) and then the angular momentum is maximized (prefactor $-\frac{1}{2} J$).

The mean-field form of the above p orbital Hamiltonian is given in Appendix \ref{append:pcase_meanfield}.

\subsection{The multi-orbital model Hamiltonian: d orbital symmetry}
\label{ss:dcase}
The on-site Coulomb interaction for cubic harmonic d orbitals can be expressed as
\begin{align}
    \nonumber
       V_{\alpha\beta,\chi\gamma} &= \frac{1}{2}\bigg(U\delta_{\alpha\chi}\delta_{\beta\gamma}\\
    \nonumber
        &+\left(J+\frac{5}{2}\Delta J\right)\left(\delta_{\alpha\gamma}\delta_{\beta\chi}+\delta_{\alpha\beta}\delta_{\gamma\chi}\right)\\
    \label{eq:Vtotoperator_dorb}  
       & -48\Delta J\sum_{abdt}\xi_{\alpha ab}\xi_{\beta bd}\xi_{\chi dt}\xi_{\gamma ta}\bigg),
\end{align}
where $\xi$ is a five-component vector of traceless symmetric $3\times 3$ transformation matrices defined as follows
\begin{eqnarray}
    \nonumber\left[\xi_{1ab}\right]&=&\left(\begin{array}{ccc}
        -\frac{1}{2\sqrt{3}} & 0 & 0\\
        0 & -\frac{1}{2\sqrt{3}} & 0\\
        0 & 0 & \frac{1}{\sqrt{3}}
    \end{array}\right),\\
    \nonumber \left[\xi_{2ab}\right]&=&\left(\begin{array}{ccc}
        0 & 0 & \frac{1}{2}\\
        0 & 0 & 0\\
        \frac{1}{2} & 0 & 0
    \end{array}\right),\\
    \nonumber \left[\xi_{3ab}\right]&=&\left(\begin{array}{ccc}
        0 & 0 & 0\\
        0 & 0 & \frac{1}{2}\\
        0 & \frac{1}{2} & 0
    \end{array}\right),\\
    \nonumber \left[\xi_{4ab}\right]&=&\left(\begin{array}{ccc}
        0 & \frac{1}{2} & 0\\
        \frac{1}{2} & 0 & 0\\
        0 & 0 & 0
    \end{array}\right),\\
    \label{eq:xi}
    \left[\xi_{5ab}\right]&=&\left(\begin{array}{ccc}
        \frac{1}{2} & 0 & 0\\
        0 & -\frac{1}{2} & 0\\
        0 & 0 & 0
    \end{array}\right).
\end{eqnarray}
We use the convention that index numbers $(1,2,3,4,5)$ correspond to the d orbitals $(3z^2-r^2,zx,yz,xy,x^2-y^2)$. See Appendix \ref{subappend:dcase} for a derivation of Eq.~(\ref{eq:xi}) using representation theory. We have chosen $U$ to be the Hartree term between the~$t_{2g}$ orbitals, $J$ to be the average of the exchange integral between the $e_g$ and $t_{2g}$ orbitals, and $\Delta J$ to be the difference between the exchange integrals for $e_g$ and $t_{2g}$. That is, $U = V_{(zx)(yz),(zx)(yz)}$, $J=\frac{1}{2}\left(V_{(zx)(yz),(yz)(zx)}+V_{(3z^{2}-r^2)(x^{2}-y^{2}),(x^{2}-y^{2})(3z^{2}-r^2)}\right)$, and $\Delta J=V_{(3z^{2}-r^2)(x^{2}-y^{2}),(x^{2}-y^{2})(3z^{2}-r^2)}-V_{(zx)(yz),(yz)(zx)}$. This choice of parameters is the same as was used by Ole\'{s} and Stollhoff\cite{oles1984}; Eq.~(\ref{eq:Vtotoperator_dorb}) generalizes their result and restores rotational invariance in orbital space. Note that, in the mean field, the on-site block of the density matrix in a cubic solid is diagonal; this will cause the mean-field version of Eq.~\ref{eq:Vtotoperator_dorb_operators}) to reduce to that of Ole\'{s} and Stollhoff\cite{oles1984}. However the presence of interfaces, vacancies or interstitials will break the cubic symmetry make the on-site blocks of the density matrix non-diagonal. Therefore the use of the complete Hamiltonian is recommended in mean-field calculations for systems that do not have cubic symmetry.

Rewriting Eq.~(\ref{eq:Vtotoperator_dorb}) in terms of rotationally invariant operators gives
\begin{align}
    \nonumber
        \hat{V}_{\mrm{tot}} &= \frac{1}{2}\bigg[\left(U-\frac{1}{2}J+5\Delta J\right):\hat{n}^{2}:-\frac{1}{2}\left(J-6\Delta J\right):\hat{{\mbf m}}^{2}:\\
    \label{eq:Vtotoperator_dorb_operators}
        &+\left(J-6\Delta J\right)\sum_{\alpha\beta}:\left(\hat{n}_{\alpha\beta}\right)^2: +\frac{2}{3}\Delta J :\hat{Q}^2:\bigg],
\end{align}
where $\hat{Q}^2$ is the on-site quadrupole operator squared. 
The quadrupole operator for a single electron is defined as
\begin{equation}
    \label{eq:quadrupole_op}
        \hat{Q}_{\mu \nu}=\frac{1}{2}\left(\hat{L}_\mu\hat{L}_\nu+\hat{L}_\nu\hat{L}_\mu\right)-\frac{1}{3}\delta_{\mu \nu}\hat{\mbf L}^2.
\end{equation}
The quadrupole operator for a system of many electrons is obtained by summing the operators for each electron. Since each term in the sum acts on the coordinates of one electron only, the operators $\hat{L}_{\mu}$ and $\hat{L}_{\nu}$ appearing in that term measure the angular momentum of one electron only, not the whole system.
More detail on the $\hat{Q}^2$ operator is included in Appendix \ref{append:quadrupole}. Eq.~(\ref{eq:Vtotoperator_dorb_operators}) is similar in form to that found in Rudzikas\cite{rudzikas1997}, the difference being that our Hamiltonian is expressed in terms of physically meaningful operators.

The mean-field form of this Hamiltonian is given in Appendix \ref{append:dcase_meanfield}.

\subsection{Comparison with the Stoner Hamiltonian}
\label{ss:stoner}
The Stoner Hamiltonian for p and d orbital atoms, as generally defined in the literature, has the following form for the on-site Coulomb interaction:\cite{duc2009}
\begin{equation}
    \label{eq:stonerHami}
        \hat{V}_{\mathrm{Stoner}} = \frac{1}{2}\left(U - \frac{1}{2} J \right):\hat{n}^{2}: - \frac{1}{4}J:\hat{m}_{z}^{2}:,
\end{equation}
where $J$, in this many-body context, is the Stoner parameter, $I$. However, in the mean field, the Stoner parameter should not be taken to be equal to $J$ due to the self-interaction error; see reference \cite{stollhoff1990stoner}. This form clearly breaks rotational symmetry in spin space, which can be restored by substituting $:\hat{m}^2_{z}:$ with $:\hat{\mbf{m}}^2:$
\begin{equation}
    \label{eq:Msq_stonerHami}
        \hat{V}_{\hat{\mbf{m}}^2\mathrm{~Stoner}} = \frac{1}{2}\left( U - \frac{1}{2} J \right):\hat{n}^{2}: - \frac{1}{4}J:\hat{\mbf{m}}^{2}:.
\end{equation}
In this paper we will compare our Hamiltonians, Eqs.~(\ref{eq:C_pcase_2}) and (\ref{eq:Vtotoperator_dorb_operators}), with Eq.~(\ref{eq:Msq_stonerHami}), which we shall call the vector or $\hat{\mbf{m}}^2$ Stoner Hamiltonian. By starting with the vector Stoner Hamiltonian and working backwards, it is possible to find the corresponding tensorial form of the Coulomb interaction:
\begin{equation}
    \label{eq:V_abcd_stoner}
        V^{\hat{\mbf{m}}^2\mathrm{~Stoner}}_{\alpha\beta,\chi\gamma} = U\delta_{\alpha\chi}\delta_{\beta\gamma}+J\delta_{\alpha\gamma}\delta_{\beta\chi}.
\end{equation}
This is very similar to the tensorial form for the p-case Coulomb interaction, Eq.~(\ref{eq:V_pcase}), except that it is missing a term, $J\delta_{\alpha\beta}\delta_{\gamma\chi}$. The lack of this term means that the vector Stoner Hamiltonian does not respect the symmetry between pairs $\alpha\chi$ and $\beta\gamma$ evident from the form of the integral in Eq.~(\ref{eq:V_abcd_rotinvar}), i.e., $V^{\hat{\mbf{m}}^2\mathrm{~Stoner}}_{\chi\beta,\alpha\gamma}\neq V^{\hat{\mbf{m}}^2\mathrm{~Stoner}}_{\alpha\beta,\chi\gamma}$. As we shall see later, this omission changes the computed results.

\section{Ground state results}
\label{s:GS}
\subsection{Simulation setup}
\label{ss:GS_setup}
Here we investigate the eigenstates of dimers using both our Hamiltonians, Eqs.~(\ref{eq:C_pcase_2}) and (\ref{eq:Vtotoperator_dorb_operators}), and the vector Stoner Hamiltonian, Eq.~(\ref{eq:Msq_stonerHami}). We employ a restricted basis with a single set of s, p and d orbitals for the s, p and d Hamiltonians respectively. We perform FCI calculations using the HANDE \cite{HANDE} computer program to find exact eigenstates of the model Hamiltonians. The Lanczos algorithm was used to calculate the lowest 40 wavefunctions for our Hamiltonian, more than sufficient to find the correct ground state wavefunction, while the full spectrum was calculated for the vector Stoner Hamiltonian.

The single particle contribution to the Hamiltonian for all models (the hopping matrix) is defined for a dimer aligned along the z-axis: in this case it is only non-zero between orbitals of the same type, whether on the same site or on different sites. The sigma hopping integral, $|t_{\sigma}|$, was set to 1 to define the energy scale. The other hopping matrix elements follow the canonical relations suggested by Andersen\cite{andersen1975}, and we use the values of Paxton and Finnis\cite{paxton2008}: $pp\sigma : pp\pi = 2:-1$ for p orbitals, and $dd\sigma : dd\pi : dd\delta = -6:4:-1$ for d orbitals.

The directionally averaged magnetic correlation between the two sites has been calculated for the ground state. If positive it shows that the atoms are ferromagnetically correlated (spins parallel on both sites) and if it is negative it shows that the atoms are antiferromagnetically correlated (spins antiparallel between sites). The correlation between components of spin projected onto a direction described by angles $\theta$ and $\phi$, in spherical coordinates, is defined as follows
\begin{equation}
    \label{eq:mag_corr_dir}
        C_{\theta\phi} = \langle :\hat{m}^1_{\theta\phi}\hat{m}^2_{\theta\phi}: \rangle,
\end{equation}
where the $::$ is the normal ordering operator, used to remove self-interaction, and
\begin{align}
    \label{eq:mag_dir}
        \hat{m}^I_{\theta\phi} &= \sum_{\alpha\xi\xi'} \sigma_{\xi\xi'}^{\theta\phi}\hat{c}^\dagger_{I\alpha,\xi}\hat{c}^{\phantom{\dagger}}_{I\alpha,\xi'},\\
    \nonumber
        \sigma_{\xi\xi'}^{\theta\phi} & = \sigma^z_{\xi\xi'}\cos\theta+\sigma^x_{\xi\xi'}\sin\theta\cos\phi+\sigma^y_{\xi\xi'}\sin\theta\sin\phi.
\end{align}
The magnetic correlation averaged over solid angle is
\begin{equation}
    \label{eq:mag_corr_avg}
        C_{\mathrm{avg}} = \frac{1}{3}\langle :\hat{\mbf{m}}_1.\hat{\mbf{m}}_2: \rangle.
\end{equation}

\subsection{Ground state: p orbital dimer}
\label{ss:GS_pcase}
\begin{figure}
    \includegraphics[width=0.5\textwidth]{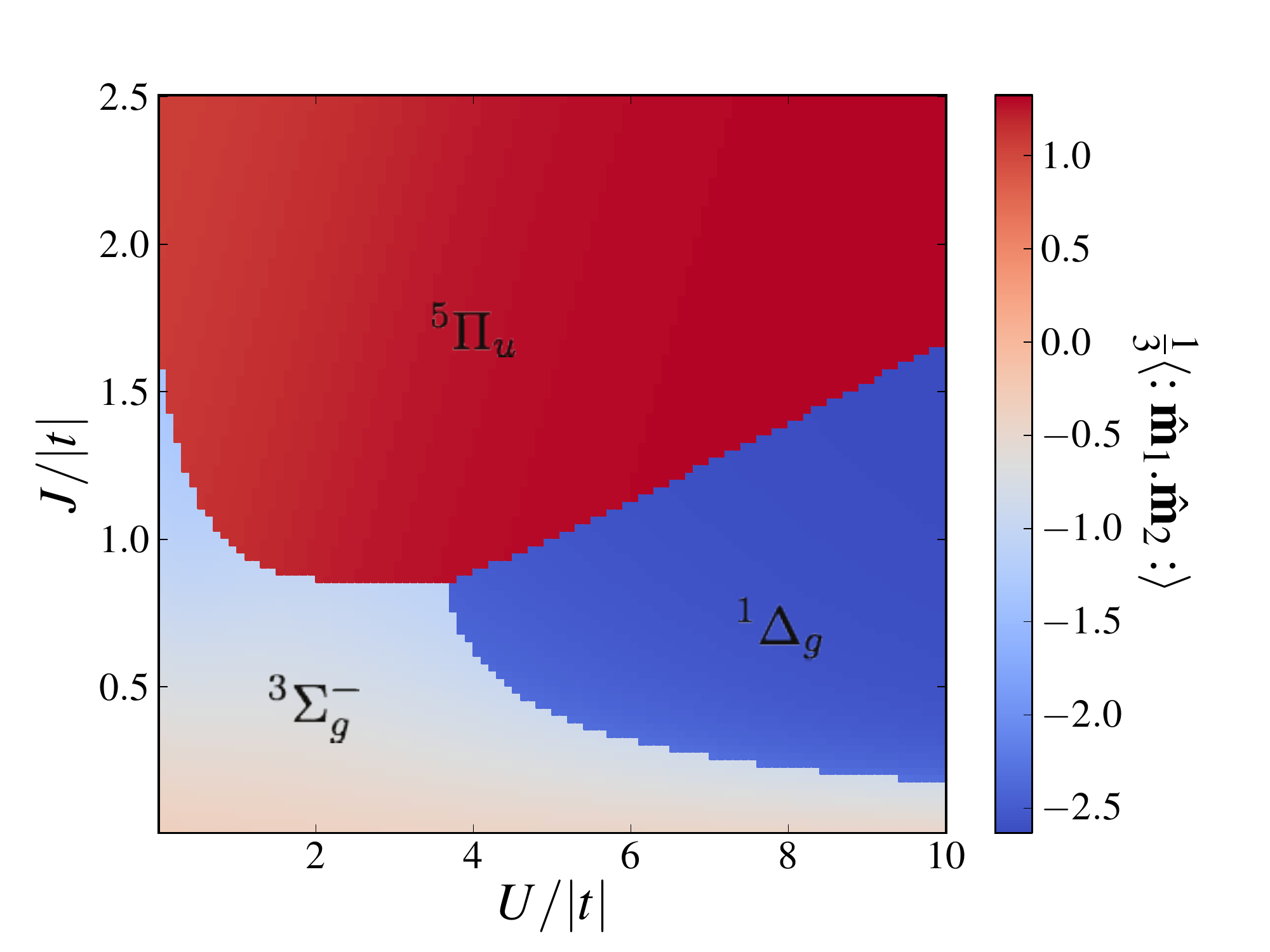}
    \includegraphics[width=0.5\textwidth]{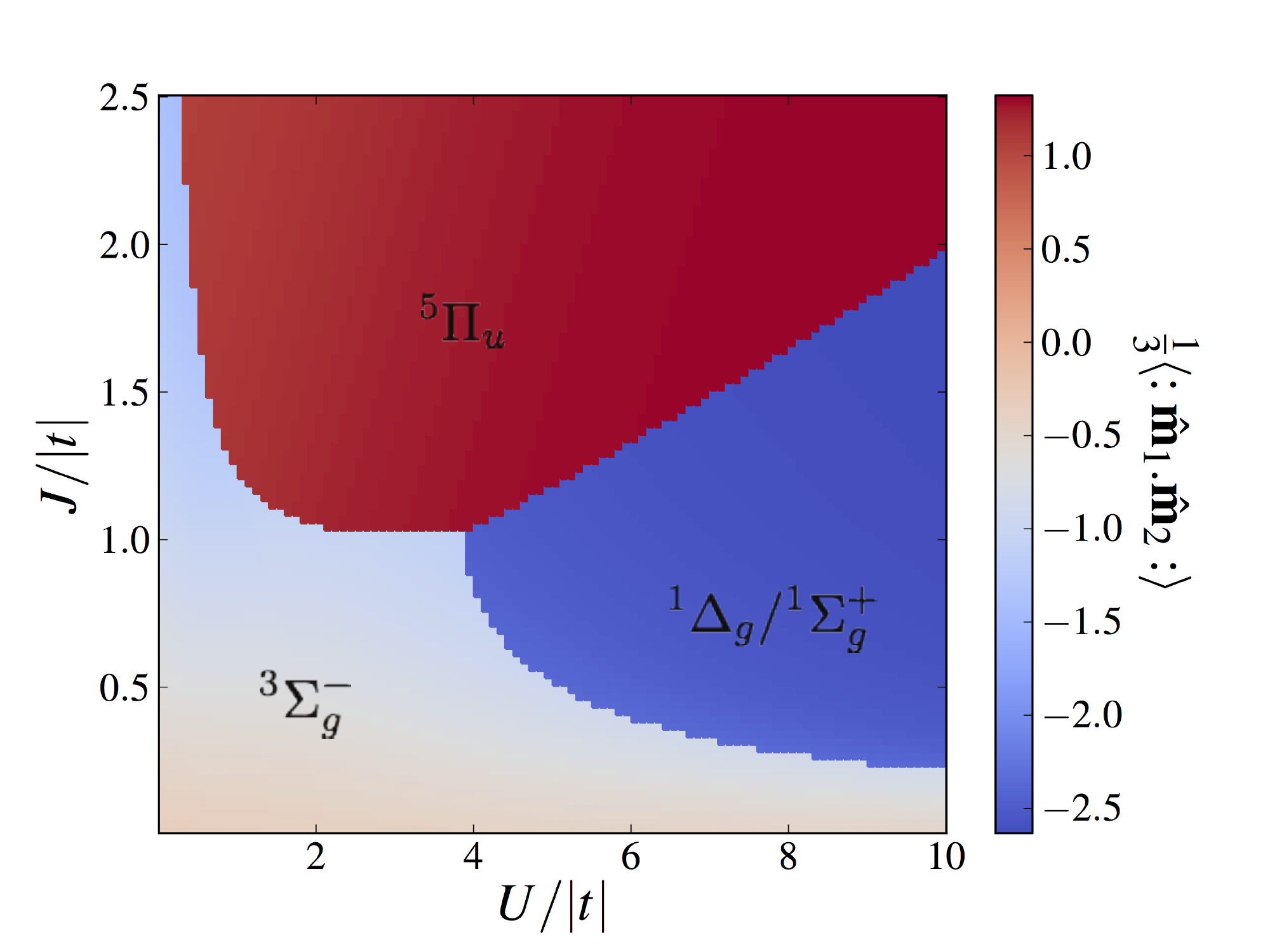}
    \caption{The magnetic correlation between two p-shell atoms, each with 2 electrons, for a large range of parameters $U/|t|$ and $J/|t|$, where $t$ is the sigma bond hopping. The different regions of the graph are labeled by the symmetry of the ground state. The top graph is generated from our Hamiltonian and the bottom from the vector Stoner Hamiltonian. The bottom graph has a region with symmetry ${}^3\Sigma^-_g$ extending a long way up the $J$ axis, which is not present in the ground state of our Hamiltonian. The bottom graph also includes a region with two degenerate states with symmetries ${}^1\Delta_g$ and ${}^1\Sigma^+_g$; this degeneracy is broken when our Hamiltonian is used.}
    \label{fig:pcase_dimer}
\end{figure}

The ground states of our Hamiltonian and the vector Stoner Hamiltonian have been found to be rather similar for 2, 6 and 10 electrons split over the p orbital dimer, but qualitatively different for 4 and 8 electrons. The magnetic correlation between the two atoms for 4 electrons is shown in Fig.~\ref{fig:pcase_dimer}. The symmetries of the wavefunctions \cite{landau1981quantum} are indicated on the graph by the notation ${}^{2S+1}\Lambda_{u/g}^{\pm}$: the $\pm$ is only used for $L_z=0$ (i.e.~$\Sigma$ states) and corresponds to the sign change after a reflection in a plane parallel to the axis of the dimer; the $u/g$ term refers to \textit{ungerade} (odd) and \textit{gerade} (even) and corresponds to a reflection through the midpoint of the dimer; $\Lambda$ is the symbol corresponding to the total value for $L_z$ (e.g.~$L_z=1$ is $\Pi$, $L_z=2$ is $\Delta$, $L_z=3$ is $\Phi$, $L_z=4$ is $\Gamma$); and $S$ is the total spin.
The differences between the ground states of our Hamiltonian and the vector Stoner Hamiltonian shown in Fig.~\ref{fig:pcase_dimer} are as follows: the ground-state wavefunction of the vector Stoner Hamiltonian has a region with ${}^3\Sigma^-_g$ symmetry extending far up the $J$ axis, which is not present for our Hamiltonian; the region with ${}^1\Delta_g$ symmetry is doubly degenerate for our Hamiltonian, as total $L_z=\pm 2$, whereas for the Stoner Hamiltonian it is triply degenerate, as it is also degenerate with the state of symmetry ${}^1\Sigma^+_g$ (this state appears at a higher energy for our Hamiltonian).

\subsection{Ground state: d orbital dimer}
\label{ss:GS_dcase}
\begin{figure}
    \includegraphics[width=0.5\textwidth]{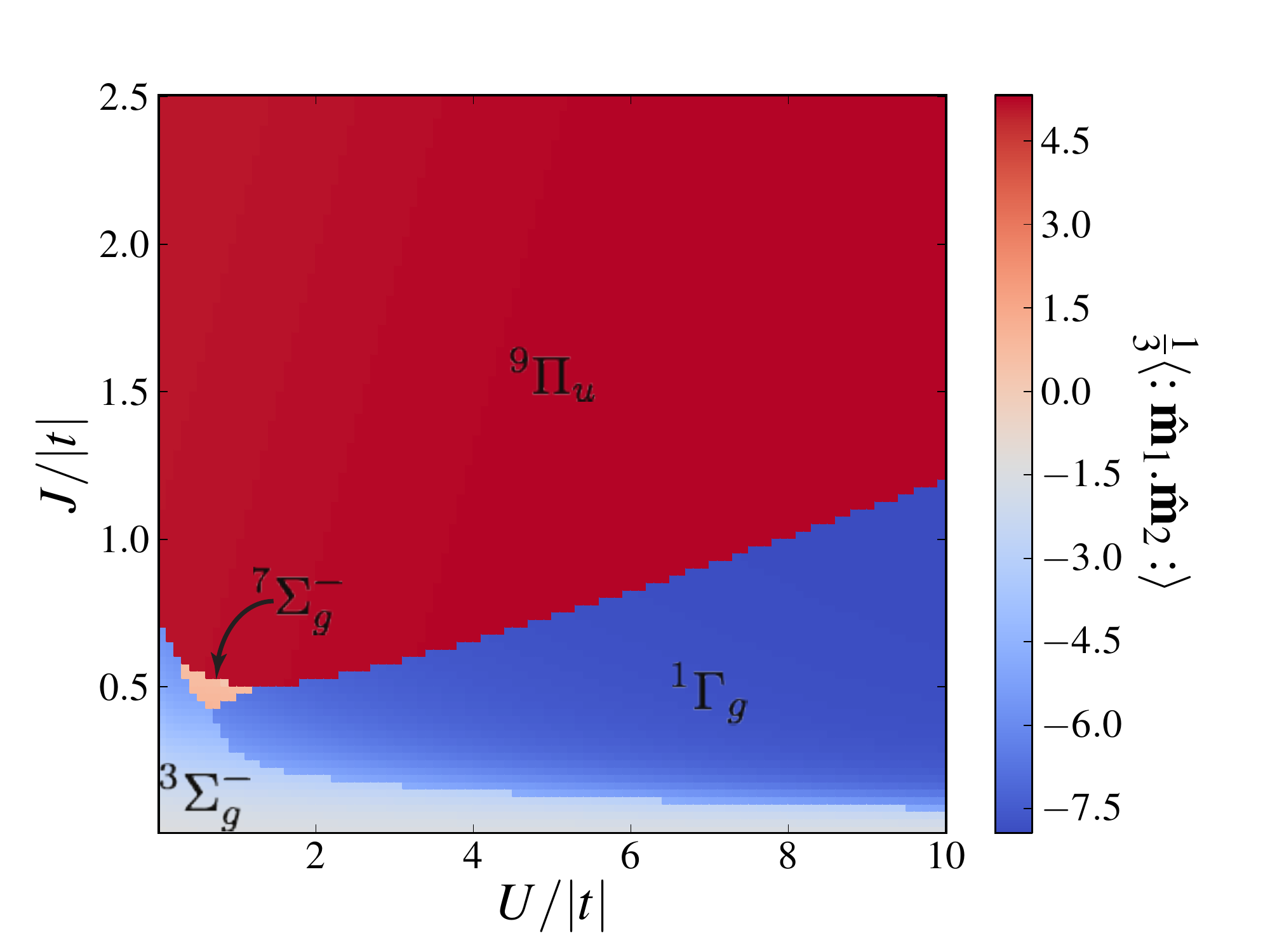}
    \includegraphics[width=0.5\textwidth]{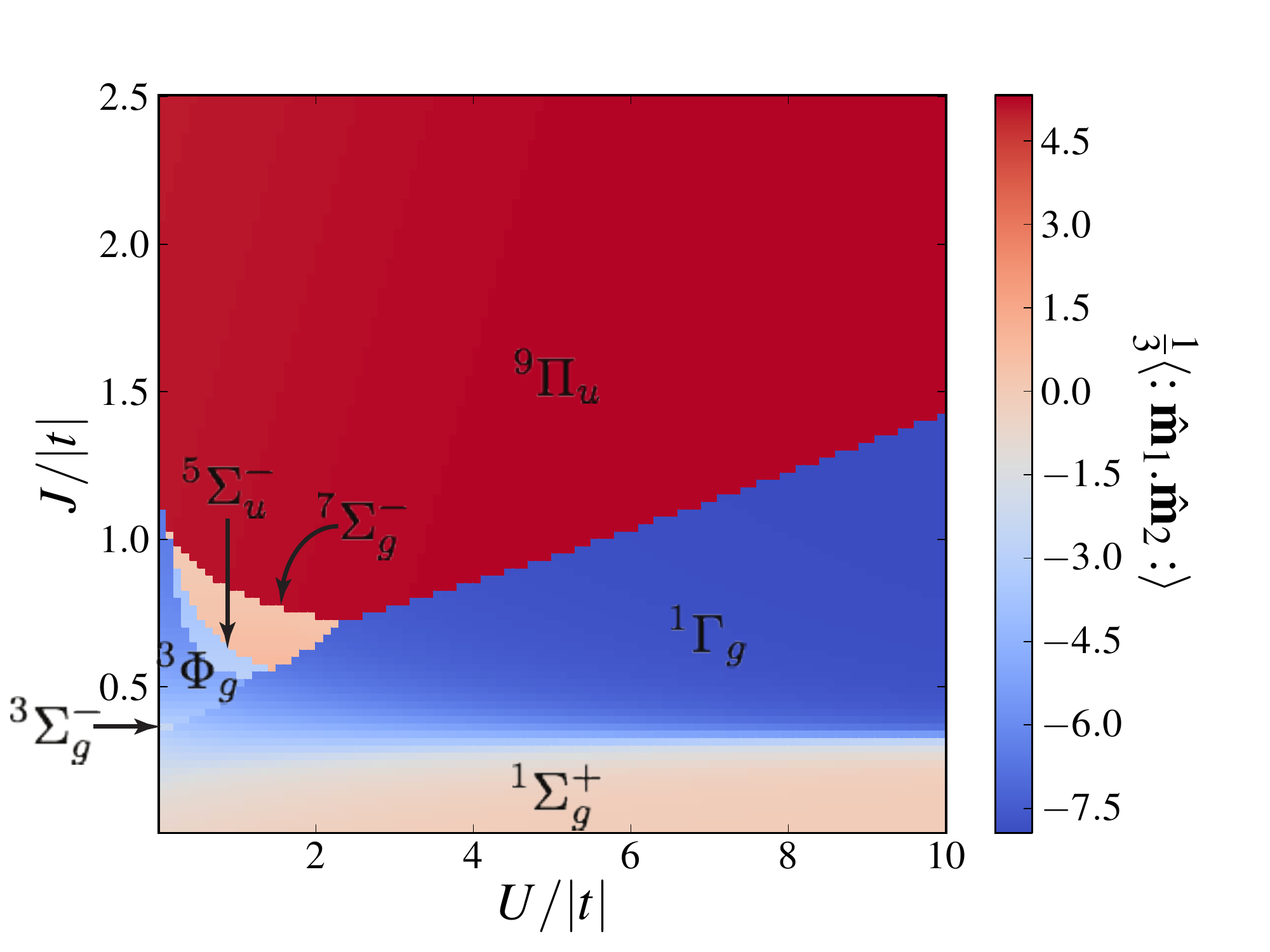}
    \includegraphics[width=0.5\textwidth]{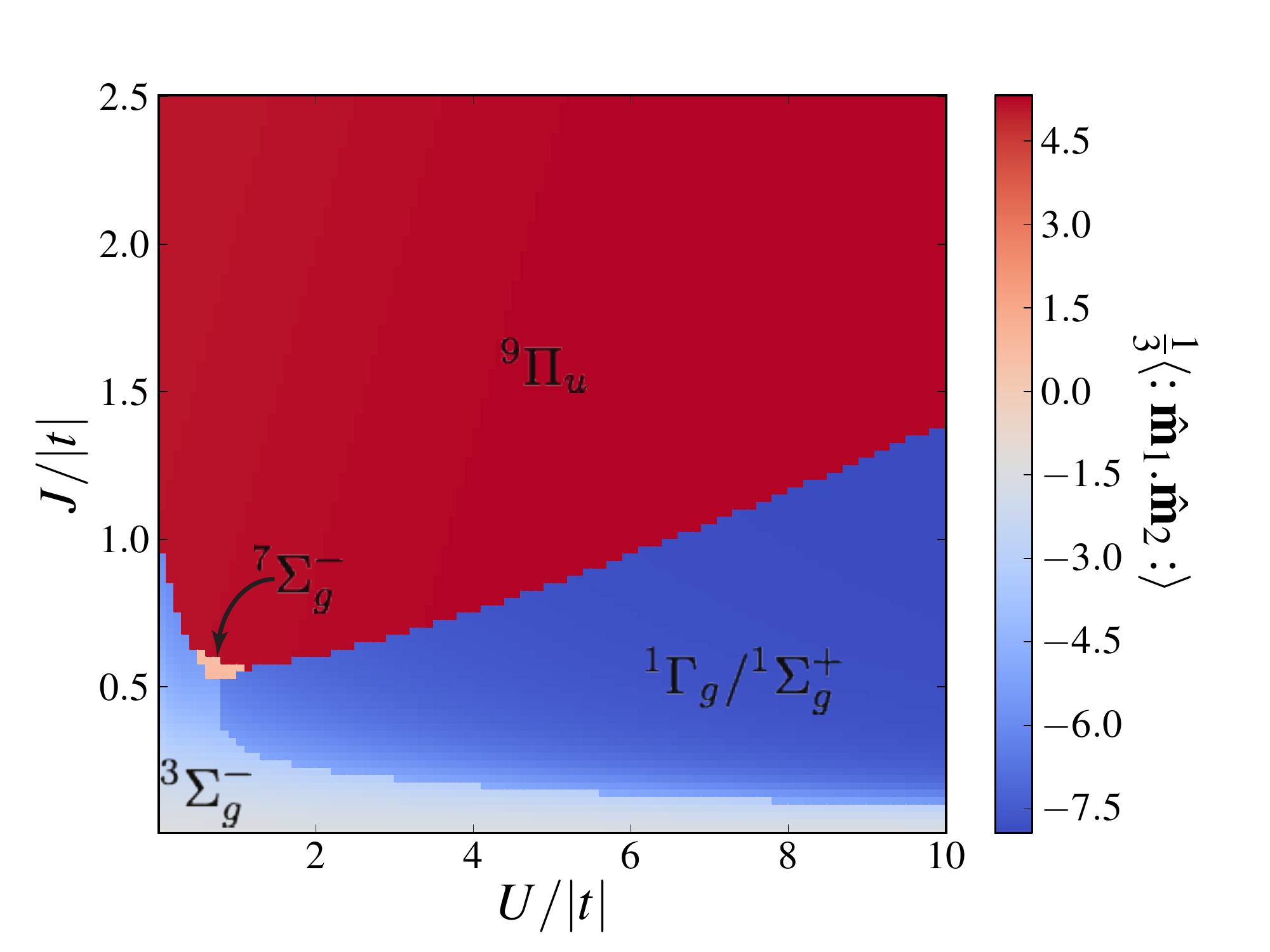}
    \caption{The magnetic correlation between two d-shell atoms, each with 6 electrons, for a large range of parameters $U/|t|$ and $J/|t|$, where $t$ is the sigma bond hopping. The different regions of the graph are labeled by the symmetry of the ground state. The top graph is generated using our d-shell Hamiltonian with $\Delta J/|t|=0.0$; the middle graph is generated using our Hamiltonian with a small value of $\Delta J/|t|=0.1$; and the bottom graph is generated using the vector Stoner Hamiltonian.}
    \label{fig:dcase_dimer}
\end{figure}

The ground states of our Hamiltonian and vector Stoner Hamiltonian have been found to be rather similar for 2, 4, 6, 14, and 18 electrons split over the d orbital dimer when $\Delta J$ is small. The simulations for 10 electrons were not carried out as they are too computationally expensive. Qualitative differences were found for 8 and 12 electrons; for an example see Fig.~\ref{fig:dcase_dimer}, which shows the magnetic correlation between two atoms in the ground state of the d-shell dimer with 6 electrons per atom.
From Fig.~\ref{fig:dcase_dimer} we see that the results for our Hamiltonian with $\Delta J/|t|=0.0$ (top graph) and with a small value of $\Delta J/|t| = 0.1$ (middle graph) are qualitatively different from those for the vector Stoner Hamiltonian (bottom graph). The vector Stoner Hamiltonian makes the ${}^1\Gamma_g$ and ${}^1\Sigma^+_g$ states degenerate, which is not the case for our Hamiltonian. The largest differences are between the $\Delta J/|t| = 0.1$ graph and the vector Stoner Hamiltonian: new regions with symmetry ${}^1\Sigma^+_g$, ${}^3\Phi_{g}$ and ${}^5\Sigma^-_g$ appear, and the region with symmetry ${}^3\Sigma^-_g$ almost disappears. This shows that the inclusion of the quadrupole term in Eq.~(\ref{eq:Vtotoperator_dorb_operators}) can make a qualitative difference to the ground state.

\section{Excited state results}
\label{s:excitations}
Differences between our Hamiltonian and the vector Stoner Hamiltonian are also observed for excited states. Here we present three examples, one demonstrating explicitly the effect of the pair hopping, one showing more general differences in the excited states through a calculation of the electronic heat capacity as a function of temperature, and one showing a difference in the spin dynamics of a collinear and a non-collinear Hamiltonian. The full spectrum of eigenstates was calculated for all of these examples.

\subsection{Pair hopping}
\label{ss:pair_hopping}
\begin{figure}
    \centering
    \includegraphics[width=0.5\textwidth]{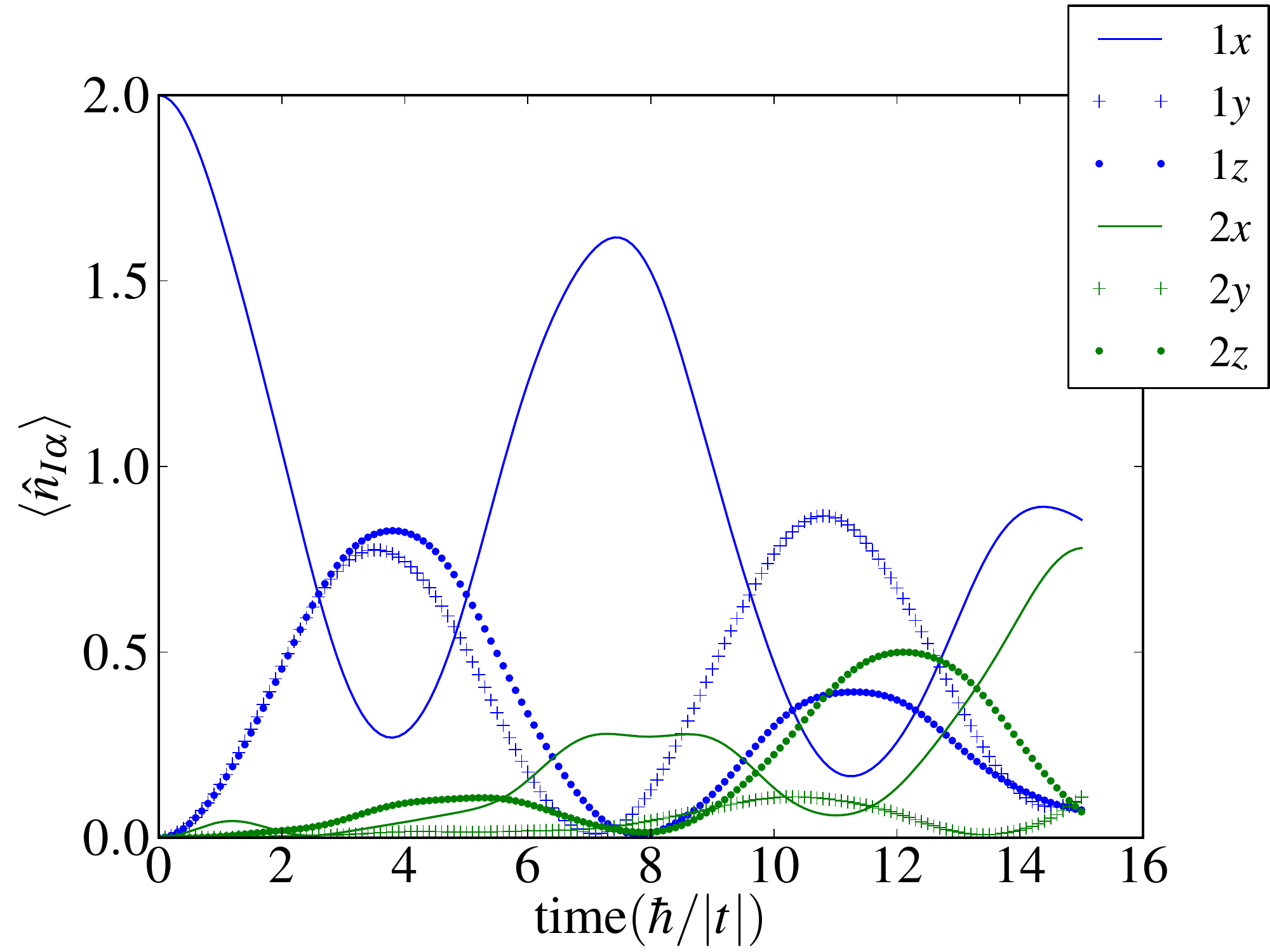}
    \includegraphics[width=0.5\textwidth]{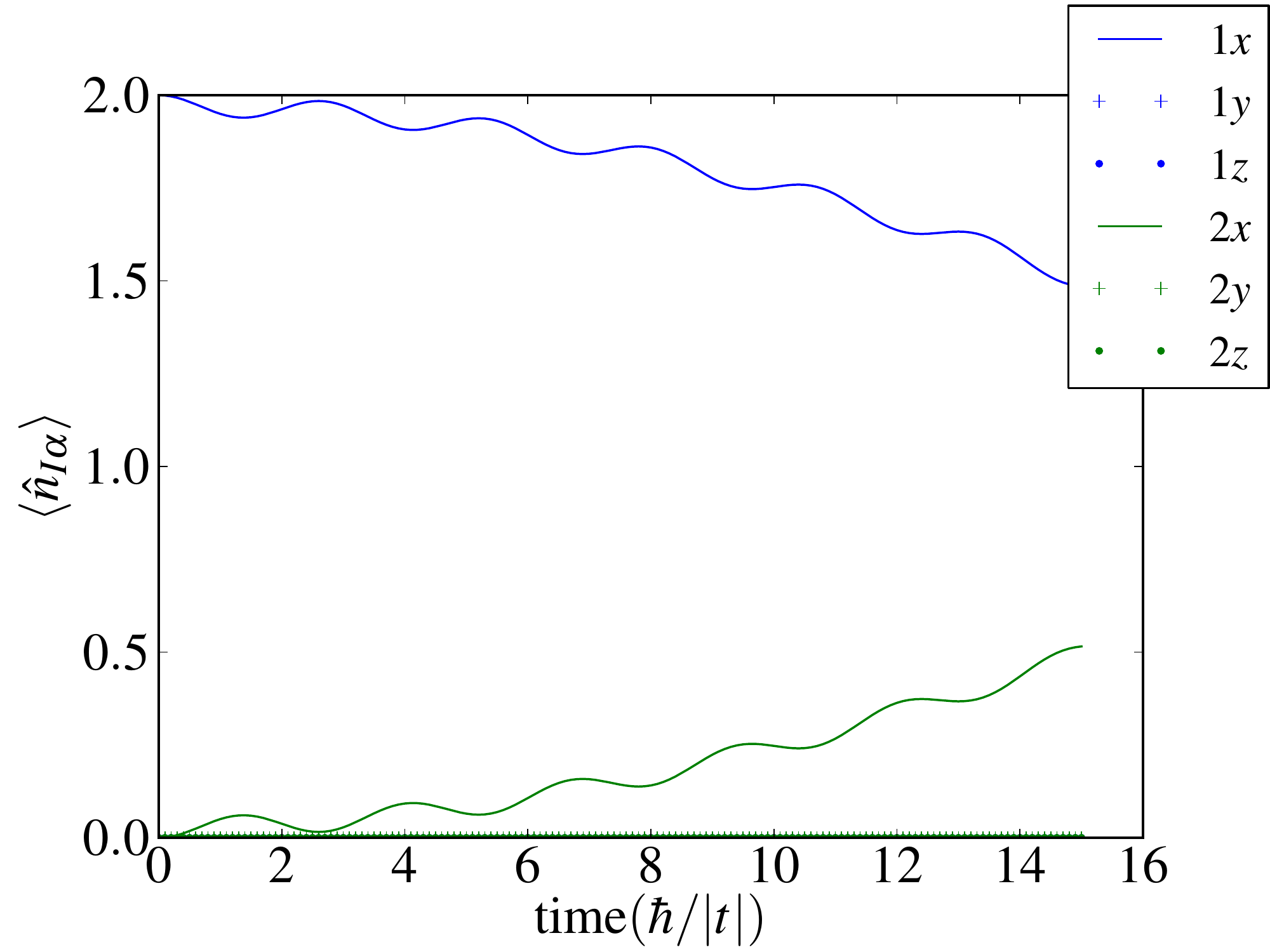}
    \caption{The time evolution under our Hamiltonian (top) and the vector Stoner Hamiltonian (bottom) of a starting state with two electrons in $1x$, the $x$ orbital on atom 1, of a p-atom dimer with $U=5.0|t|$ and $J=0.7|t|$. We see that the two electrons in $1x$ are able to hop into $1y$ and $1z$ when the wavefunction is evolved using our Hamiltonian but not when it is evolved using the vector Stoner Hamiltonian.}
    \label{fig:pair_hopping}
\end{figure}

The pair hopping term, $J\sum_{\alpha\beta\sigma\sigma'}\hat{c}^{\dagger}_{\alpha,\sigma}\hat{c}^{\dagger}_{\alpha,\sigma'}\hat{c}_{\beta,\sigma'}\hat{c}_{\beta,\sigma} = J \sum_{\alpha\beta}:\left(\hat{n}_{\alpha\beta}\right)^2:$, is found in both our p and d orbital Hamiltonians, but is absent from the vector Stoner Hamiltonian.
To demonstrate the effect of the pair hopping term we initialize the wavefunction in a state with two electrons in the $x$ orbital on site 1 of a p-atom dimer. 
Figure \ref{fig:pair_hopping} shows the evolution of the wavefunction in time using our Hamiltonian and the vector Stoner Hamiltonian, with $U=5.0|t|$ and $J=0.7|t|$.
For our Hamiltonian the pair of electrons in orbital $x$ on atom 1 hops to the $y$ and $z$ orbitals on atom 1 (mediated by the pair hopping term) as well as between the two atoms (mediated by the single electron hopping term). The result is that there is a finite probability of finding the pair of electrons in any of the $x$, $y$ and $z$ orbitals on either atom. However, for the vector Stoner Hamiltonian, the two electrons in the $x$ orbital on site 1 are unable to hop into the $y$ and $z$ orbitals on atom 1; they are only able to hop to the $x$ orbital on atom 2. This means tha there is no possibility of observing the pair of electrons in the $y$ or $z$ orbitals on either atom.

\subsection{Heat capacity}
\label{ss:heat_capacity}
\begin{figure}
    \centering
    \includegraphics[width=0.5\textwidth]{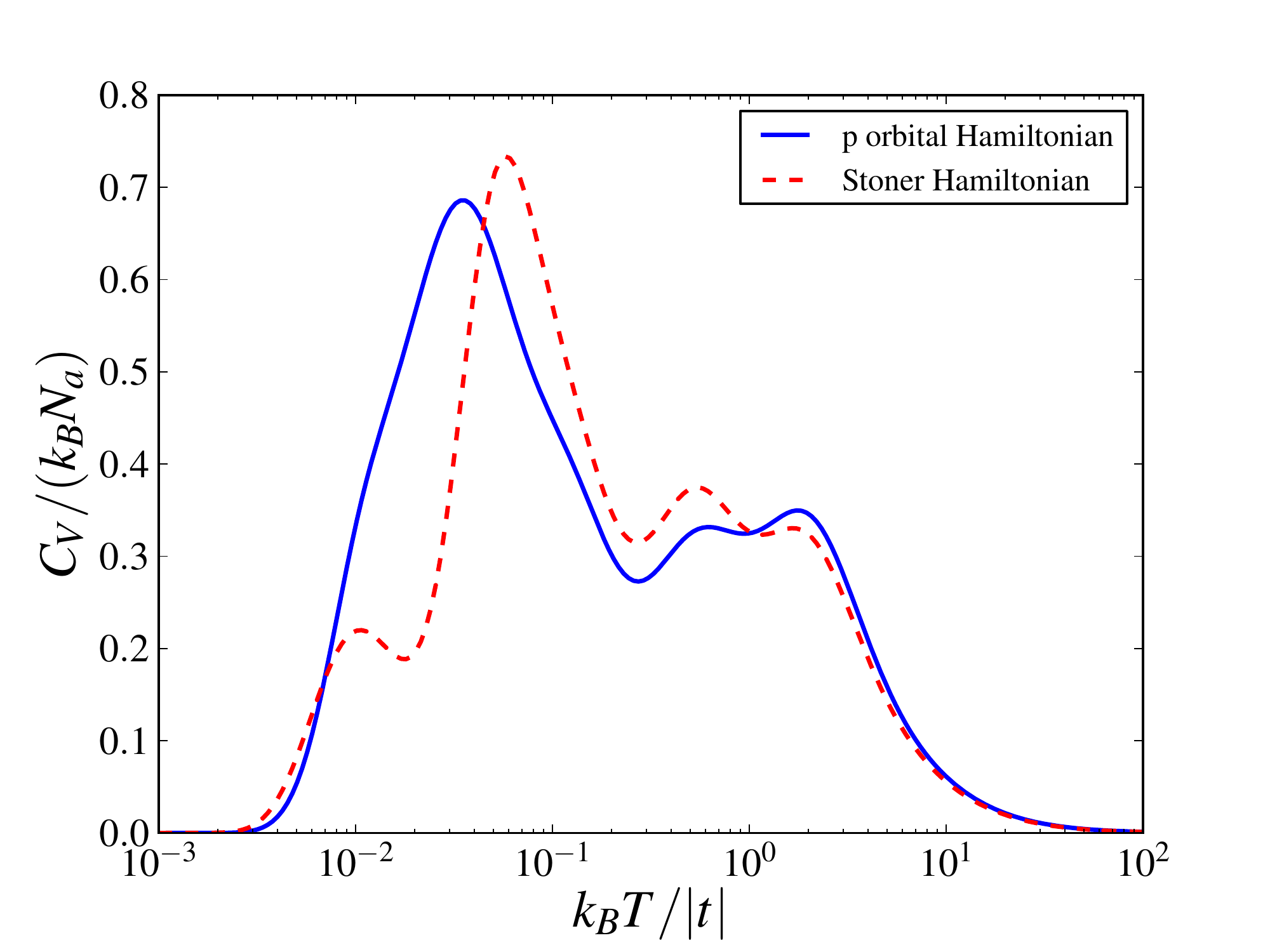}
    \caption{The electronic heat capacity per atom of a dimer with four electrons split over two p-shell atoms with $U=5.0|t|$ and $J=0.7|t|$. The solid line is generated by solving our Hamiltonian exactly and the dashed line by solving the vector Stoner Hamiltonian exactly.}
    \label{fig:heat_capacity}
\end{figure}

The heat capacity is calculated from
\begin{equation}
    \label{eq:heat_capacity}
        \frac{C_V}{k_B N_a} = \frac{1}{N_a(k_BT)^2}\left(\langle E^2(T)\rangle-\langle E(T)\rangle^2\right),
\end{equation}
where
\begin{align}
    \label{eq:internal_energy}
        \langle E^n(T)\rangle & = \frac{\sum_{i}\eps^n_i \exp(-\frac{\eps_i}{k_B T})}{\sum_i \exp(-\frac{\eps_i}{k_B T})},
\end{align}
$T$ is the temperature, $k_B$ is Boltzmann's constant, $\eps_i$ is a many-electron energy eigenvalue, and $N_a$ is the number of atoms. The result for both Hamiltonians as a function of temperature for four electrons split over two p-shell atoms with $U=5.0|t|$ and $J=0.7|t|$ is shown in Figure \ref{fig:heat_capacity}.
There is a qualitative difference between the results for our Hamiltonian (the solid line) and the vector Stoner Hamiltonian (dashed line). The energies of the eigenstates of our Hamiltonian are more spread out than those of the Stoner Hamiltonian. This is due to the inclusion of the pair hopping term, which causes states with on-site paired electrons to rise in energy (by $\sim J/|t|$). The unphysical reduction of the heat capacity to zero as the temperature tends to infinity is a consequence of the use of a restricted basis set.

\subsection{Spin dynamics}
\label{ss:spin_dynamics}
Here we show how the collinear Stoner Hamiltonian can give rise to unphysical spin dynamics. We consider an $S_z=+1$ triplet with both electrons on one of the two p orbital atoms, $|\Psi\rangle = |T_{+1}\rangle$. Written as linear combinations of two-electron Slater determinants, the three states in the triplet are
\begin{align}
    \nonumber
        |T_{+1}\rangle = & \frac{1}{\sqrt{2}}\left(-|1x\uparrow 1y\uparrow\rangle + |2x\uparrow 2y\uparrow\rangle\right),\\
    \nonumber
        |T_{-1}\rangle = & \frac{1}{\sqrt{2}}\left(-|1x\downarrow 1y\downarrow\rangle + |2x\downarrow 2y\downarrow\rangle\right),\\
    \nonumber
        |T_0\rangle = & -\frac{1}{\sqrt{2}}\left(|1x\downarrow 1y\uparrow\rangle + |1x\uparrow 1y\downarrow\rangle \right)\\
    \label{eq:triplet}
        & + \frac{1}{\sqrt{2}}\left( |2x\downarrow 2y\uparrow\rangle + |2x\uparrow 2y\downarrow\rangle \right).
\end{align}
These states are simultaneously eigenstates of the collinear Stoner Hamiltonian, the vector Stoner Hamiltonian, and our p-case Hamiltonian. 
They are degenerate, with eigenvalue $U-J$, for both our p-case Hamiltonian and for the vector Stoner Hamiltonian.
For the collinear Stoner Hamiltonian the states $|T_{+1}\rangle$ and $|T_{-1}\rangle$ have eigenvalue $U-J$ whereas state $|T_0\rangle$ has eigenvalue $U$.
This means that the degeneracy of this triplet is broken in the collinear Stoner Hamiltonian. We now rotate $|\Psi\rangle$ in spin space so that the spins are aligned with the $x$-axis, i.e. $S_x=+1$,
\begin{align}
    \label{eq:psi_rot}
        |\Psi^\mathrm{rot}\rangle = & \frac{1}{2}\left(|T_{+1}\rangle+|T_{-1}\rangle\right)+\frac{1}{\sqrt{2}}|T_0\rangle.
\end{align} 
$|\Psi^\mathrm{rot}\rangle$ is still an eigenstate, with eigenvalue $U-J$, of both the vector Stoner Hamiltonian and our Hamiltonian, but it is no longer an eigenstate of the collinear Stoner Hamiltonian. The wavefunction $|\Psi^{\mathrm{rot}}\rangle$ evolves in time as
\begin{align}
    \label{eq:psi_rot_time}
        |\Psi^{\mathrm{tot}}(\tau)\rangle &= e^{-\frac{i\hat{H}\tau}{\hbar}}|\Psi^{\mathrm{tot}}(0)\rangle,\\
    \nonumber
         & = \left\{\begin{array}{c}
             \underbrace{e^{-\frac{i(U-J)\tau}{\hbar}}\left(\frac{1}{2}\left(|T_{+1}\rangle+|T_{-1}\rangle\right) +\frac{e^{-\frac{iJ\tau}{\hbar}}}{\sqrt{2}}|T_0\rangle\right)}_\mathrm{collinear~Stoner~Hamiltonian},\\
             \underbrace{e^{-\frac{i(U-J)\tau}{\hbar}}\left(\frac{1}{2}\left(|T_{+1}\rangle+|T_{-1}\rangle\right)+\frac{1}{\sqrt{2}}|T_0\rangle\right)}_{\hat{\mathbf{m}}^2~\mathrm{Stoner~and~our~Hamiltonian}},
         \end{array}\right.
\end{align}
where $\tau$ is time. If we take the expectation value of the magnetic moment on each site using the collinear Stoner Hamiltonian, we find
\begin{align}
    \nonumber
        \langle\Psi^\mathrm{rot}(\tau)|\hat{m}_{1x}|\Psi^\mathrm{rot}(\tau)\rangle &= \langle\Psi^\mathrm{rot}(\tau)|\hat{m}_{2x}|\Psi^\mathrm{rot}(\tau)\rangle = \cos\left(\frac{J\tau}{\hbar}\right),\\
    \nonumber
        \langle\Psi^\mathrm{rot}(\tau)|\hat{m}_{1y}|\Psi^\mathrm{rot}(\tau)\rangle &= \langle\Psi^\mathrm{rot}(\tau)|\hat{m}_{2y}|\Psi^\mathrm{rot}(\tau)\rangle = 0,\\
    \label{eq:collinear_spin_dynamics}
        \langle\Psi^\mathrm{rot}(\tau)|\hat{m}_{1z}|\Psi^\mathrm{rot}(\tau)\rangle &= \langle\Psi^\mathrm{rot}(\tau)|\hat{m}_{2z}|\Psi^\mathrm{rot}(\tau)\rangle = 0.
\end{align}
In contrast, using the vector Stoner Hamiltonian or our p-shell Hamiltonian, the expectation value of the magnetic moment is independent of time, equal to $(1,0,0)$ in vector format.
This demonstrates that calculations of spin dynamics using the collinear Stoner Hamiltonian can give rise to unphysical oscillations of the magnetic moments.

\section{Conclusion}
\label{s:conc}
We have established the correct form of the multi-orbital model Hamiltonian with on-site Coulomb interactions for atoms with valence shells of s, p and d orbitals. The methodology used may be extended to atoms with f and g shells, and to atoms with valence orbitals of several different angular momenta. The results presented show that there are important differences between our p- and d-shell Hamiltonians and the vector Stoner Hamiltonian. The vector Stoner Hamiltonian misses both the pair hopping term, which is present in our p and d orbital Hamiltonians, and the quadrupole term, which is present in our d orbital Hamiltonian. The pair hopping term pushes states with pairs of electrons up in energy, whereas the magnetism term pulls states with local magnetic moments down in energy. The pair hopping term has the largest effect on the ground state for p orbitals with 2 and 4 electrons per atom and for d orbitals with 4 and 6 electrons per atom, for $J/|t|<2$. 
This is because the number of possible determinants with paired electrons on site is large for these filling factors and the low lying states can be separated based on this.
At values of $J/|t|>2$ the magnetism becomes the dominant effect upon the selection of the ground state and the difference between the ground state of our Hamiltonian and that of the vector Stoner Hamiltonian becomes small. However, the differences are rather more pronounced for the excited states. This is evidenced by the hopping of pairs of electrons between orbitals on the same site, which is allowed by our Hamiltonian but not by the vector Stoner Hamiltonian,  and in differences in the electronic heat capacity as a function of temperature. We also find clear evidence that the collinear Stoner Hamiltonian is inappropriate for use in describing spin dynamics as it breaks rotational symmetry in spin space. We would expect similar problems when using collinear time dependent DFT simulations to model spin dynamics.

\section*{Acknowledgements}
The authors would like to thank Toby Wiseman for help with the representation theory and Dimitri Vvedensky, David M.~Edwards, Cyrille Barreteau and Daniel Spanjaard for stimulating discussions.
M.~E.~A.~Coury was supported through a studentship in the Centre for Doctoral Training on Theory and Simulation of Materials at Imperial College funded by EPSRC under grant number EP/G036888/1.
This work was part-funded by the EuroFusion Consortium, and has received funding from Euratom research and training programme 2014–2018 under Grant Agreement No. 633053, and funding from the RCUK Energy Programme (Grant No. EP/I501045). The views and opinions expressed herein do not necessarily reflect those of the European Commission.
The authors acknowledge support from the Thomas Young Centre under grant TYC-101

\appendix
\section{Mean-field Hamiltonians}
\label{append:meanfield}

\subsection{The one band Hubbard model: s orbital symmetry}
\label{append:scase_meanfield}
Application of the mean-field approximation \cite{bruus2004} to the on-site Coulomb interaction part of the Hubbard Hamiltonian with s orbital symmetry, Eq.~(\ref{eq:C_op_num_scase}), yields the following, 
\begin{align}
    \label{eq:scase_num_meanfield}
        \hat{V}_{\mathrm{MF}} = & U_0 \Bigg(\langle\hat{n}\rangle\hat{n}-\sum_{\sigma\zeta}\langle\hat{c}^\dagger_{\sigma}\hat{c}^{\phantom{\dagger}}_{\zeta}\rangle\hat{c}^\dagger_{\zeta}\hat{c}^{\phantom{\dagger}}_{\sigma}\Bigg),
\end{align}
for which the total energy is
\begin{align}
    \label{eq:scase_num_energy}
        E^{\mathrm{Coulomb}}_{\mathrm{MF}} = & \frac{1}{2}U_0\left(\langle\hat{n}\rangle^2-\sum_{\sigma\zeta}\langle\hat{c}^\dagger_{\sigma}\hat{c}^{\phantom{\dagger}}_{\zeta}\rangle\langle\hat{c}^\dagger_{\zeta}\hat{c}^{\phantom{\dagger}}_{\sigma}\rangle\right),
\end{align}
where the double counting correction has been included.
Equivalently, applying the mean-field approximation to Eq.~(\ref{eq:C_op_mag_scase}) yields the following,
\begin{align}
    \label{eq:scase_mag_meanfield}
        \hat{V}_{\mathrm{MF}} = & \frac{1}{3}U_0\Bigg(2\langle\hat{n}\rangle\hat{n}-\langle\hat{\mathbf{m}}\rangle.\hat{\mathbf{m}}-\sum_{\sigma\zeta}\langle\hat{c}^\dagger_{\sigma}\hat{c}^{\phantom{\dagger}}_{\zeta}\rangle\hat{c}^\dagger_{\zeta}\hat{c}^{\phantom{\dagger}}_{\sigma}\Bigg),
\end{align}
for which the total energy is
\begin{align}
    \label{eq:scase_mag_energy}
        E^{\mathrm{Coulomb}}_{\mathrm{MF}} = & \frac{1}{6}U_0\left(2\langle\hat{n}\rangle^2-\langle\hat{\mathbf{m}}\rangle^2-\sum_{\sigma\zeta}\langle\hat{c}^\dagger_{\sigma}\hat{c}^{\phantom{\dagger}}_{\zeta}\rangle\langle\hat{c}^\dagger_{\zeta}\hat{c}^{\phantom{\dagger}}_{\sigma}\rangle\right),
\end{align}
where again the double counting correction has been included.

\subsection{The multi-orbital model Hamiltonian: p orbital symmetry}
\label{append:pcase_meanfield}
Application of the mean-field approximation \cite{bruus2004} to the
model Hamiltonian with p orbital symmetry, Eq.~(\ref{eq:C_pcase_2}), yields the following, 
\begin{align}
    \nonumber   
        \hat{V}_{\mathrm{MF}} &= \left(U - \frac{J}{2}\right)\langle\hat{n}\rangle\hat{n}-\frac{J}{2}\langle\hat{\bm m}\rangle.\hat{\bm m}\\
    \nonumber
        & +\sum_{\alpha\beta}J\bigg(\langle\hat{n}^{\phantom{\dagger}}_{\alpha\beta}\rangle\hat{n}^{\phantom{\dagger}}_{\beta\alpha}+\langle\hat{n}^{\phantom{\dagger}}_{\alpha\beta}\rangle\hat{n}^{\phantom{\dagger}}_{\alpha\beta}\bigg)\\
    \label{eq:Vtot_porb_MF}
        & -\sum_{\alpha\beta\sigma\zeta}\Bigg(U\langle\hat{c}^\dagger_{\alpha\sigma}\hat{c}^{\phantom{\dagger}}_{\beta\zeta}\rangle\hat{c}^{\dagger}_{\beta\zeta}\hat{c}^{\phantom{\dagger}}_{\alpha\sigma}+J\langle\hat{c}^{\dagger}_{\alpha\sigma}\hat{c}^{\phantom{\dagger}}_{\beta\zeta}\rangle\hat{c}^{\dagger}_{\alpha\zeta}\hat{c}^{\phantom{\dagger}}_{\beta\sigma}\Bigg),
\end{align}
for which the total energy is
\begin{align}
    \nonumber
        E^{\mathrm{Coulomb}}_{\mathrm{MF}} = & \frac{1}{2}\left(U - \frac{J}{2}\right)\langle\hat{n}\rangle^2-\frac{J}{4}\langle\hat{\bm m}\rangle^2\\
    \nonumber
        &+\sum_{\alpha\beta}\frac{J}{2}\left(\langle\hat{n}^{\phantom{\dagger}}_{\alpha\beta}\rangle\langle\hat{n}^{\phantom{\dagger}}_{\beta\alpha}\rangle+\langle\hat{n}^{\phantom{\dagger}}_{\alpha\beta}\rangle^2\right)\\
    \nonumber
        &-\sum_{\alpha\beta\sigma\zeta}\bigg(\frac{U}{2}\langle\hat{c}^\dagger_{\alpha\sigma}\hat{c}^{\phantom{\dagger}}_{\beta\zeta}\rangle\langle\hat{c}^{\dagger}_{\beta\zeta}\hat{c}^{\phantom{\dagger}}_{\alpha\sigma}\rangle\\
    \label{eq:pcase_MF_energy}
        &+\frac{J}{2}\langle\hat{c}^{\dagger}_{\alpha\sigma}\hat{c}^{\phantom{\dagger}}_{\beta\zeta}\rangle\langle\hat{c}^{\dagger}_{\alpha\zeta}\hat{c}^{\phantom{\dagger}}_{\beta\sigma}\rangle\bigg),
\end{align}
where the double counting correction has been included.

\subsection{The multi-orbital model Hamiltonian: d-orbital symmetry}
\label{append:dcase_meanfield}
Application of the mean-field approximation \cite{bruus2004} to the model Hamiltonian with d-orbital symmetry, Eq.~(\ref{eq:Vtotoperator_dorb_operators}), yields the following,
\begin{align}
    \nonumber
        \hat{V}_{\mathrm{MF}} &= \left(U-\frac{1}{2}J+2\Delta J\right)\langle\hat{n}\rangle\hat{n}-\frac{J}{2}\langle\hat{\mathbf{m}}\rangle.\hat{\mathbf{m}}\\
    \nonumber
        &+\left(J-6\Delta J\right)\sum_{\alpha\beta}\bigg(\langle\hat{n}^{\phantom{\dagger}}_{\alpha\beta}\rangle\hat{n}^{\phantom{\dagger}}_{\beta\alpha}+\langle\hat{n}^{\phantom{\dagger}}_{\alpha\beta}\rangle\hat{n}^{\phantom{\dagger}}_{\alpha\beta}\bigg)\\
    \nonumber
        &+\frac{2}{3}\Delta J\sum_{\mu\nu}\langle\hat{Q}^{\phantom{\dagger}}_{\mu\nu}\rangle\hat{Q}^{\phantom{\dagger}}_{\nu\mu}\\
    \nonumber
        &-\sum_{\alpha\beta\sigma\zeta}\Bigg(U\langle\hat{c}^\dagger_{\alpha\sigma}\hat{c}^{\phantom{\dagger}}_{\beta\zeta}\rangle\hat{c}^\dagger_{\beta\zeta}\hat{c}^{\phantom{\dagger}}_{\alpha\sigma}+J\langle\hat{c}^\dagger_{\alpha\sigma}\hat{c}^{\phantom{\dagger}}_{\beta\zeta}\rangle\hat{c}^\dagger_{\alpha\zeta}\hat{c}^{\phantom{\dagger}}_{\beta\sigma}\Bigg)\\
    \label{eq:dcase_MF}
        &+48 \Delta J \sum_{\alpha\beta\gamma\chi\sigma\zeta}\sum_{stuv}\xi^{\phantom{\dagger}}_{\alpha s t}\xi^{\phantom{\dagger}}_{\beta t u}\xi^{\phantom{\dagger}}_{\chi u v}\xi^{\phantom{\dagger}}_{\gamma v s}\langle\hat{c}^\dagger_{\alpha\sigma}\hat{c}^{\phantom{\dagger}}_{\gamma\zeta}\rangle\hat{c}^\dagger_{\beta\zeta}\hat{c}^{\phantom{\dagger}}_{\chi\sigma},
\end{align}
for which the total energy is
\begin{align}
    \nonumber
        E^{\mathrm{Coulomb}}_{\mathrm{MF}} = & \frac{1}{2}\left(U-\frac{1}{2}J+2\Delta J\right)\langle\hat{n}\rangle^2-\frac{J}{4}\langle\hat{\bm m}\rangle^2\\
    \nonumber
        &+\sum_{\alpha\beta}\frac{J-6\Delta J}{2}\left(\langle\hat{n}^{\phantom{\dagger}}_{\alpha\beta}\rangle\langle\hat{n}^{\phantom{\dagger}}_{\beta\alpha}\rangle+\langle\hat{n}^{\phantom{\dagger}}_{\alpha\beta}\rangle^2\right)\\
    \nonumber
        &+\frac{1}{3}\Delta J \sum_{\mu\nu}\langle\hat{Q}^{\phantom{\dagger}}_{\mu\nu}\rangle\langle\hat{Q}^{\phantom{\dagger}}_{\nu\mu}\rangle\\
    \nonumber
        &-\sum_{\alpha\beta\sigma\zeta}\bigg(\frac{U}{2}\langle\hat{c}^\dagger_{\alpha\sigma}\hat{c}^{\phantom{\dagger}}_{\beta\zeta}\rangle\langle\hat{c}^{\dagger}_{\beta\zeta}\hat{c}^{\phantom{\dagger}}_{\alpha\sigma}\rangle\\
    \nonumber
        &+\frac{J}{2}\langle\hat{c}^{\dagger}_{\alpha\sigma}\hat{c}^{\phantom{\dagger}}_{\beta\zeta}\rangle\langle\hat{c}^{\dagger}_{\alpha\zeta}\hat{c}^{\phantom{\dagger}}_{\beta\sigma}\rangle\bigg)\\
    \nonumber
        &+24 \Delta J \sum_{\alpha\beta\gamma\chi\sigma\zeta}\sum_{stuv}\xi^{\phantom{\dagger}}_{\alpha s t}\xi^{\phantom{\dagger}}_{\beta t u}\xi^{\phantom{\dagger}}_{\chi u v}\xi^{\phantom{\dagger}}_{\gamma v s}\\
    \label{eq:dcase_MF_energy}
        &\times\langle\hat{c}^\dagger_{\alpha\sigma}\hat{c}^{\phantom{\dagger}}_{\gamma\zeta}\rangle\langle\hat{c}^\dagger_{\beta\zeta}\hat{c}^{\phantom{\dagger}}_{\chi\sigma}\rangle,
\end{align}
where the double counting correction has been included.

\section{Representation theory}
\label{append:representation}
\subsection{p orbital symmetry}
\label{subappend:pcase}
As a precursor to our treatment of the d case in Appendix \ref{subappend:dcase}, we use representation theory\cite{fulton1991} to derive Eq.~(\ref{eq:V_pcase}) for cubic harmonic p orbitals. First we define the following irreducible objects: ``0'' is a scalar, ``1'' is a three-dimensional vector, ``2'' is a $3\times3$ symmetric traceless second-rank tensor, ``3'' is a third-rank tensor consisting of three ``2''s, and ``4'' is a fourth-rank tensor consisting of three ``3''s. The Coulomb interaction $V_{\alpha\beta,\gamma\delta}$ transforms as a tensor product of four ``1''s, $1\otimes1\otimes1\otimes1$, but we know also that it must be isotropic and that from the above irreducible objects only the ``0'' is isotropic.

Objects such as $1\otimes1$, represented as $A_{ij}$, may be expanded just as in the addition of angular momentum:
\begin{equation}
    \label{eq:reptheory1x1}
        1\otimes1  = 0 \oplus 1 \oplus 2,
\end{equation}
where ``0'' is a scalar, $s=\sum_{ij}A_{ij}\delta_{ij}$, ``1'' is a vector, $v_i = \sum_{jk}A_{jk}\epsilon_{ijk}$, and ``2'' is a traceless symmetric tensor, $M_{ij} = \frac{1}{2}(A_{ij}+A_{ji})-\frac{1}{3}\delta_{ij}(\sum_{kl}A_{kl}\delta_{kl})$.
The reader may be more familiar with combinations of angular momentum. In the language of angular momentum, the transformation properties of a tensor product of two objects of angular momentum with $l=1$ are described by a $9\times9$ matrix, which is block diagonalisable into a $1\times1$ matrix (which describes the transformations of an $l=0$ object), a $3\times3$ matrix (which describes the transformations of an $l=1$ object), and a $5\times5$ matrix (which describes the transformations of an $l=2$ object).
In the angular momentum representation the matrix elements are complex.
Here we are using cubic harmonics and the matrix elements are real.
Similarly $1\otimes1\otimes1\otimes1$ may be expanded out as:
\begin{align}
    \nonumber
        1\otimes1\otimes1&\otimes1  = (1\otimes1)\otimes(1\otimes1), \\
    \nonumber
        & = (0 \oplus 1 \oplus 2)\otimes(0 \oplus 1 \oplus 2), \\
    \nonumber
         & = (0\otimes0) \oplus (0\otimes1) \oplus (0\otimes2) \oplus (1\otimes0) \oplus (1\otimes1) \\
    \label{eq:reptheory1x1x1x1}
        & \oplus (1\otimes2) \oplus (2\otimes0) \oplus (2\otimes1) \oplus (2\otimes2).
\end{align}
The only isotropic, ``0'', objects arise from $0\otimes0$, $1\otimes1$ and $2\otimes2$. 
A general isotropic three-dimensional fourth-rank tensor, $T_{ijkl}$, therefore has three independent scalars that are found from symmetric and antisymmetric contractions of $T_{ijkl}$. 
We are interested in the form of the p orbital on-site Coulomb interactions, $V_{\alpha\beta,\chi\gamma}$, which have an additional symmetry, such that they remain unchanged under exchange of $\alpha$ and $\chi$ and under exchange of $\beta$ and $\gamma$.
We therefore only require the symmetric scalars that arise from the $0\otimes0$ and $2\otimes2$ contractions to describe $V_{\alpha\beta,\chi\gamma}$:
\begin{equation}
    \label{eq:V_porb_tensor}
        V_{\alpha\beta,\chi\gamma} = s_0\delta_{\alpha\chi}\delta_{\beta\gamma} + s_2\left(\delta_{\alpha\gamma}\delta_{\beta\chi}+\delta_{\alpha\beta}\delta_{\chi\gamma}\right),
\end{equation}
which is symmetric under exchange of $\alpha$ and $\chi$ or $\beta$ and $\gamma$; where 
\begin{align}
    \nonumber
        s_0 & = \sum_{\chi\gamma}V_{\alpha\beta,\chi\gamma}\delta_{\alpha\chi}\delta_{\beta\gamma},\\
    \label{eq:pcase_contractions}
        s_2 & = \sum_{\chi\gamma}V_{\alpha\beta,\chi\gamma}\delta_{\alpha\gamma}\delta_{\beta\chi} = \sum_{\chi\beta}V_{\alpha\beta,\chi\gamma}\delta_{\alpha\beta}\delta_{\chi\gamma}.
\end{align}
This is equivalent to Eq.~(\ref{eq:V_pcase}), where $s_0 = U$ and $s_2 = J$.

\subsection{d orbital symmetry}
\label{subappend:dcase}
To find the isotropic five-dimensional fourth-rank tensor that describes the on-site Coulomb interactions for d orbital symmetry, we find it convenient to map from a five-dimensional fourth-rank tensor to a three-dimensional eighth-rank tensor. We do this by replacing the five-dimensional vector of d orbitals by the three-dimensional traceless symmetric $B$ matrix of the d orbitals: 
\begin{equation}
    \label{eq:Bmat}
    \resizebox{0.438\textwidth}{!}{${\mbf B}=\left(\begin{array}{ccc}
        \frac{2}{3}x^{2}-\frac{1}{3}\left(y^{2}+z^{2}\right) & xy & xz\\
        xy & \frac{2}{3}y^{2}-\frac{1}{3}\left(x^{2}+z^{2}\right) & yz\\
        xz & yz & \frac{2}{3}z^{2}-\frac{1}{3}\left(x^{2}+y^{2}\right)
    \end{array}\right).$}
\end{equation}
We refer to the elements of ${\mbf B}$ as $B_{ab}$, and use the subscripts $a$, $b$, $c$, $d$, $s$, $t$, $u$ and $v$ as the indices for irreducibles of rank 2 or higher. The transformation between the irreducible $B_{ab}$ and the cubic harmonic d orbitals is
\begin{equation}
    \label{eq:dorb_trans_irrep_cub}
        \phi_{\alpha\sigma}({\mbf r})=N_d\sum_{ab}\xi_{\alpha ab}B_{ab}\frac{1}{r^2} R_d(r)S(\sigma),
\end{equation}
where $\xi$ is a traceless, symmetric transformation matrix, defined in Eq.~(\ref{eq:xi}), $N_d = \frac{1}{2}\sqrt{\frac{15}{\pi}}$ is a normalisation factor, $R_d$ is the radial function for a cubic d orbital, and $S$ is the spin function. The orbital indices $\alpha$, $\beta$, $\chi$ and $\gamma$ run over the five independent d orbitals, whereas the irreducible indices $a$, $b$, $c$, $d$, $s$, $t$, $u$ and $v$ run over the three Cartesian directions. The mapping between the five-dimensional fourth-rank tensor and the three-dimensional eighth-rank tensor is as follows
\begin{align}
    \nonumber
        V_{\alpha\beta,\chi\gamma} & = \int \dd {\mbf r}_1 \dd {\mbf r}_2 \frac{\phi_{\alpha\sigma}^{*}({\mbf r}_{1})\phi_{\beta\sigma'}^{*}({\mbf r}_{2})\phi_{\chi\sigma}^{\phantom{*}}({\mbf r}_{1})\phi_{\gamma\sigma'}^{\phantom{*}}({\mbf r}_{2})}{|{\mbf r}_1-{\mbf r}_2|}, \\
    \nonumber
         & = \int \dd {\mbf r}_1 \dd {\mbf r}_2 N_d^4 |R_d(r_1)/r_1^2|^2|R_d(r_2)/r_2^2|^2\\
    \nonumber
        & \times\frac{\sum_{abcd}\sum_{stuv}\xi^*_{\alpha ab}B^*_{ab}\xi^*_{\beta cd}B^*_{cd} \xi^{\phantom{*}}_{\chi st}B^{\phantom{*}}_{st}\xi^{\phantom{*}}_{\gamma uv} B^{\phantom{*}}_{uv}}{|{\mbf r}_1-{\mbf r}_2|}, \\
    \label{eq:V_dorb_irrep}
         & = \sum_{abcd}\sum_{stuv}\xi_{\alpha ab}\xi_{\beta cd}\xi_{\chi st}\xi_{\gamma uv} V_{ab,cd,st,uv},
\end{align}
where this equation defines the isotropic three-dimensional eighth-rank tensor for the on-site Coulomb integrals, $V_{ab,cd,st,uv}$, and we have dropped the complex conjugates on the $\xi$ matrices as they are real. The $B$ matrix is a three-dimensional traceless symmetric ``2'' and thus contains 5 independent terms. As we show below, it is possible to represent the isotropic three-dimensional eighth-rank tensor as a list of quadruple Kronecker deltas, with 5 independent parameters in general and 3 independent parameters for the on-site Coulomb interactions. The isotropic three-dimensional eighth-rank tensor is then converted back into an isotropic five-dimensional fourth-rank tensor. For an independent verification of the form of the isotropic three-dimensional eighth-rank tensor see Ref.~\onlinecite{kearsley1975}.

The power of representation theory is that one can follow an analogous procedure for shells of higher angular momentum, using a ``3'' to represent the seven f orbitals, a ``4'' to represent the nine g orbitals, and so on.  One can also construct interaction Hamiltonians for atoms with important valence orbitals of several different angular momenta.  

We now return to the case of a d shell and explain the procedure used to find the general isotropic Coulomb Hamiltonian in more detail. We start by representing the isotropic three-dimensional eighth-rank tensor, $V_{ab,cd,st,uv}$, as a $2\otimes2\otimes2\otimes2$. Proceeding as we did for the p case, we write
\begin{align}
    \nonumber
        2\otimes2\otimes2\otimes2&=(2\otimes2)\otimes(2\otimes2),\\
    \nonumber
        2\otimes2&=0\oplus 1\oplus 2\oplus 3\oplus 4,\\
    \nonumber
        \implies2\otimes2\otimes2\otimes2&=(0\oplus 1\oplus 2\oplus 3\oplus 4)\\
    \label{eq:2x2x2x2}
        &\otimes(0\oplus 1\oplus 2\oplus 3\oplus 4).
\end{align}
The terms in Eq.~(\ref{eq:2x2x2x2}) that can generate a rank 0 (a scalar) are  $0\otimes0$, $1\otimes1$, $2\otimes2$, $3\otimes3$ and $4\otimes4$. This implies that a general isotropic three-dimensional eighth-rank tensor is defined by five independent parameters. We know that, for $V_{\alpha\beta,\chi\gamma}$, there exists a symmetry between the pairs $\alpha\chi$ and $\beta\gamma$. It follows that $V_{ab,cd,st,uv}$ has a symmetry between the pairs $(ab)(st)$ and $(cd)(uv)$. We therefore require only the even contractions, $0\otimes0$, $2\otimes2$ and $4\otimes4$, reducing the five parameters to three \footnote{By continuing this argument, one finds that there 4 parameters are required to describe the on-site Coulomb interactions for an f shell (arising from $0\otimes0$, $2\otimes2$, $4\otimes4$ and $6\otimes6$) and 5 parameters for a g shell (arising from $0\otimes0$, $2\otimes2$, $4\otimes4$, $6\otimes6$ and $8\otimes8$), respectively. This trend continues with increasing angular momentum.}; this is the same number proposed in Chapter 4 of Ref.~\onlinecite{griffith1961}.  To get the scalars from $V_{ab,cd,st,uv}$ we have to contract it. There are six possible contractions of $V_{ab,cd,st,uv}$, outlined in Table \ref{tab:nodecontractions}, of which only five are independent. Note that one does not contract within the indices $ab$, $cd$, $st$ or $uv$ because $B_{ab}$ is traceless, hence $\sum_{ab}\delta_{ab}B_{ab}=0$.
    \begin{table*}[h]
        $\begin{array}{cccc}
        \multirow{6}{*}{\includegraphics[scale=0.4]{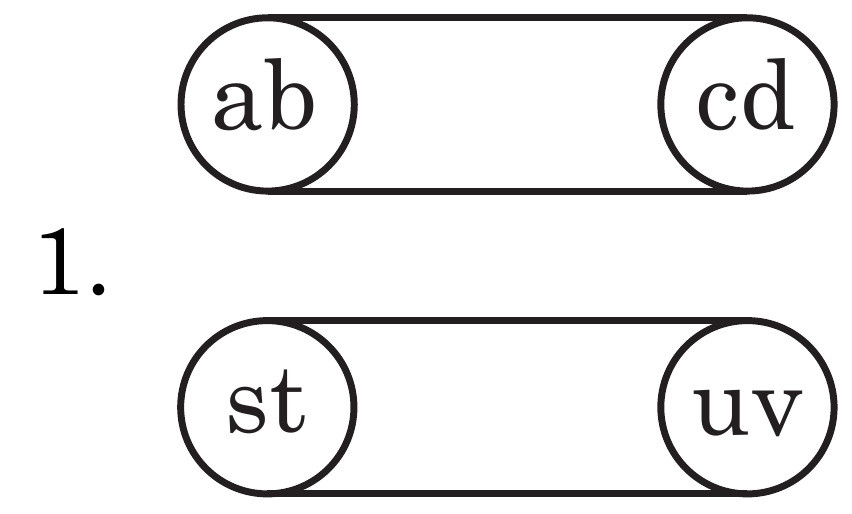}} & & &\\
        & & & \\
        & \Delta^1_{ab,cd,st,uv} & = & \delta_{ad} \delta_{bc} \delta_{sv} \delta_{tu}+\delta_{ac} \delta_{bd} \delta_{sv} \delta_{tu}+\delta_{ad} \delta_{bc} \delta_{su} \delta_{tv}+\delta_{ac} \delta_{bd} \delta_{su} \delta_{tv} \\
        & & & \\
        & & & \\
        & & & \\
        \multirow{6}{*}{\includegraphics[scale=0.4]{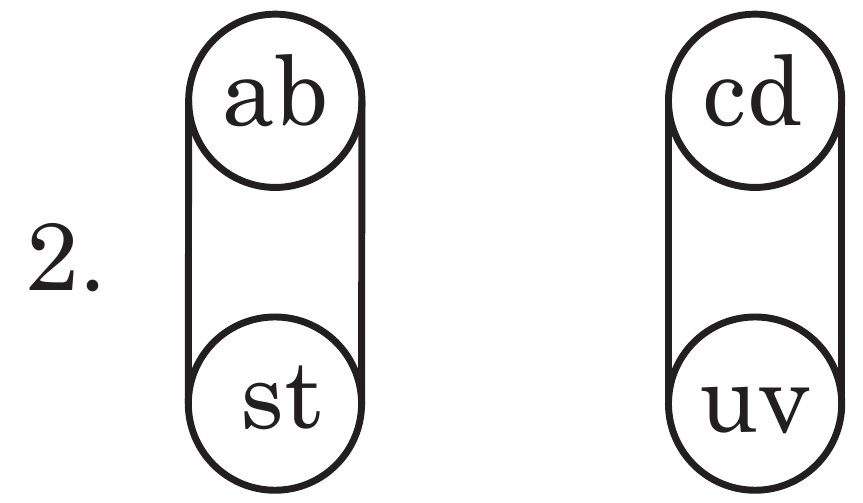}} & & & \\
        & & & \\
        & \Delta^2_{ab,cd,st,uv} & = & \delta_{at} \delta_{bs} \delta_{cv} \delta_{du}+\delta_{as} \delta_{bt} \delta_{cv} \delta_{du}+\delta_{at} \delta_{bs} \delta_{cu} \delta_{dv}+\delta_{as} \delta_{bt} \delta_{cu} \delta_{dv} \\ 
        & & & \\
        & & & \\
        & & & \\
        \multirow{6}{*}{\includegraphics[scale=0.4]{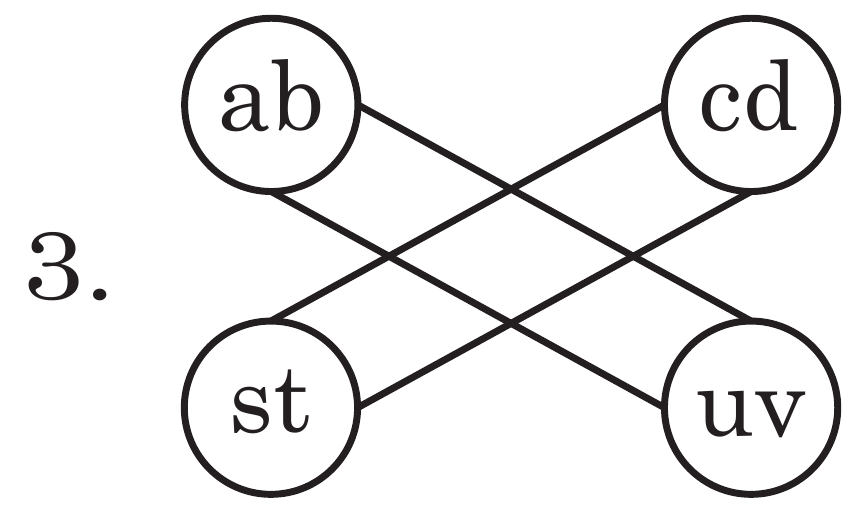}} & & & \\
        & & & \\
        & \Delta^3_{ab,cd,st,uv} & = & \delta_{av} \delta_{bu} \delta_{ct} \delta_{ds}+\delta_{au} \delta_{bv} \delta_{ct} \delta_{ds}+\delta_{av} \delta_{bu} \delta_{cs} \delta_{dt}+\delta_{au} \delta_{bv} \delta_{cs} \delta_{dt} \\
        & & & \\
        & & & \\
        & & & \\
        \multirow{6}{*}{\includegraphics[scale=0.4]{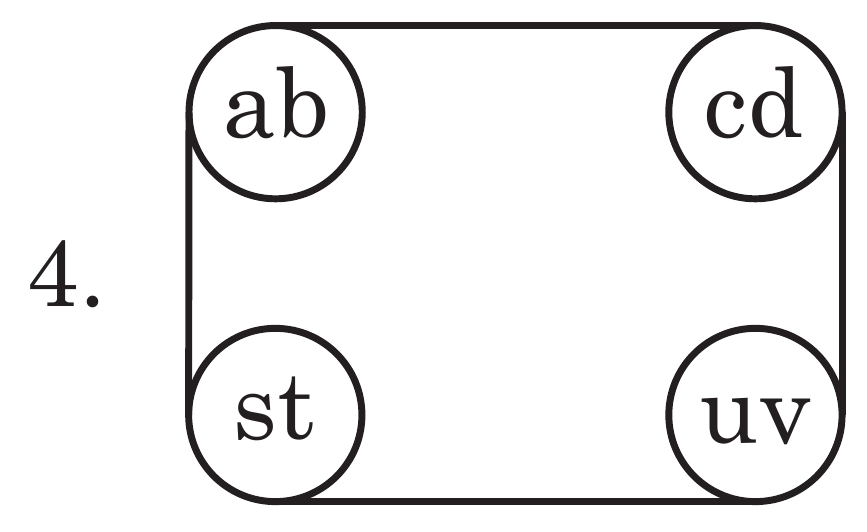}} & & & \\
        & \Delta^4_{ab,cd,st,uv} & = & \delta_{at} \delta_{bd} \delta_{cv} \delta_{su}+\delta_{ad} \delta_{bt} \delta_{cv} \delta_{su}+\delta_{at} \delta_{bc} \delta_{dv} \delta_{su}+\delta_{ac} \delta_{bt} \delta_{dv} \delta_{su} +\\
         & & & \delta_{at} \delta_{bd} \delta_{cu} \delta_{sv}+\delta_{ad} \delta_{bt} \delta_{cu} \delta_{sv}+\delta_{at} \delta_{bc} \delta_{du} \delta_{sv}+\delta_{ac} \delta_{bt} \delta_{du} \delta_{sv}+ \\
         & & & \delta_{as} \delta_{bd} \delta_{cv} \delta_{tu}+\delta_{ad} \delta_{bs} \delta_{cv} \delta_{tu}+\delta_{as} \delta_{bc} \delta_{dv} \delta_{tu}+\delta_{ac} \delta_{bs} \delta_{dv} \delta_{tu}+ \\
         & & & \delta_{as} \delta_{bd} \delta_{cu} \delta_{tv}+\delta_{ad} \delta_{bs} \delta_{cu} \delta_{tv}+\delta_{as} \delta_{bc} \delta_{du} \delta_{tv}+\delta_{ac} \delta_{bs} \delta_{du} \delta_{tv} \\
         & & & \\
         \multirow{6}{*}{\includegraphics[scale=0.4]{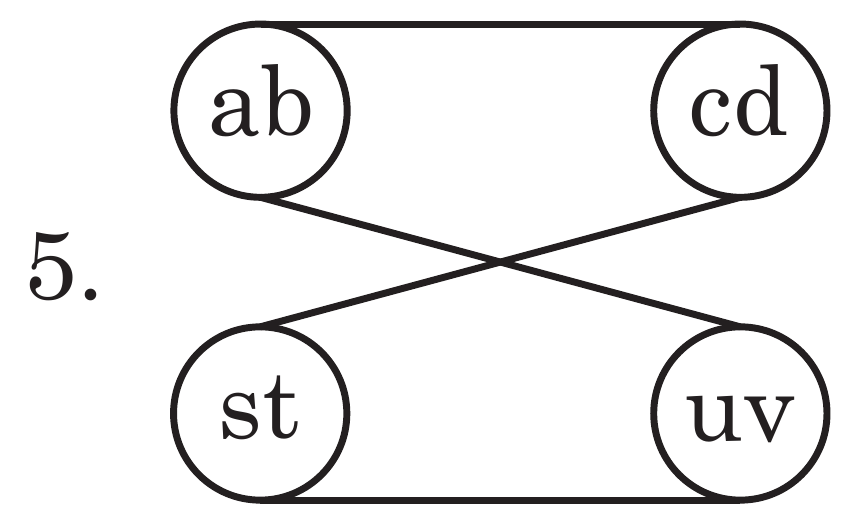}} & & & \\
         & \Delta^5_{ab,cd,st,uv} & = & \delta_{av} \delta_{bd} \delta_{ct} \delta_{su}+\delta_{ad} \delta_{bv} \delta_{ct} \delta_{su}+\delta_{av} \delta_{bc} \delta_{dt} \delta_{su}+\delta_{ac} \delta_{bv} \delta_{dt} \delta_{su}+ \\
         & & &\delta_{au} \delta_{bd} \delta_{ct} \delta_{sv}+\delta_{ad} \delta_{bu} \delta_{ct} \delta_{sv}+\delta_{au} \delta_{bc} \delta_{dt} \delta_{sv}+\delta_{ac} \delta_{bu} \delta_{dt} \delta_{sv}+ \\
         & & & \delta_{av} \delta_{bd} \delta_{cs} \delta_{tu}+\delta_{ad} \delta_{bv} \delta_{cs} \delta_{tu}+\delta_{av} \delta_{bc} \delta_{ds} \delta_{tu}+\delta_{ac} \delta_{bv} \delta_{ds} \delta_{tu}+ \\
         & & & \delta_{au} \delta_{bd} \delta_{cs} \delta_{tv}+\delta_{ad} \delta_{bu} \delta_{cs} \delta_{tv}+\delta_{au} \delta_{bc} \delta_{ds} \delta_{tv}+\delta_{ac} \delta_{bu} \delta_{ds} \delta_{tv} \\
         & & & \\
         \multirow{6}{*}{\includegraphics[scale=0.4]{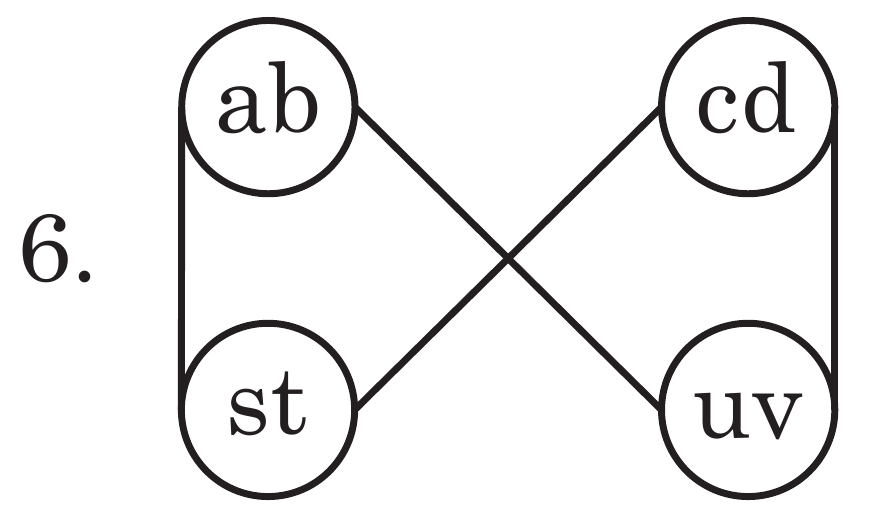}} & & & \\
         & \Delta^6_{ab,cd,st,uv} & = & \delta_{av} \delta_{bt} \delta_{cu} \delta_{ds}+\delta_{at} \delta_{bv} \delta_{cu} \delta_{ds}+\delta_{au} \delta_{bt} \delta_{cv} \delta_{ds}+\delta_{at} \delta_{bu} \delta_{cv} \delta_{ds}+ \\
         & & &\delta_{av} \delta_{bs} \delta_{cu} \delta_{dt}+\delta_{as} \delta_{bv} \delta_{cu} \delta_{dt}+\delta_{au} \delta_{bs} \delta_{cv} \delta_{dt}+\delta_{as} \delta_{bu} \delta_{cv} \delta_{dt}+  \\
         & & & \delta_{av} \delta_{bt} \delta_{cs} \delta_{du}+\delta_{at} \delta_{bv} \delta_{cs} \delta_{du}+\delta_{av} \delta_{bs} \delta_{ct} \delta_{du}+\delta_{as} \delta_{bv} \delta_{ct} \delta_{du}+ \\
         & & & \delta_{au} \delta_{bt} \delta_{cs} \delta_{dv}+\delta_{at} \delta_{bu} \delta_{cs} \delta_{dv}+\delta_{au} \delta_{bs} \delta_{ct} \delta_{dv}+\delta_{as} \delta_{bu} \delta_{ct} \delta_{dv} \\
         & & &
        \end{array}$
        \caption{Table to show the node contractions for a d shell. Every contraction is represented both diagrammatically and mathematically.}
        \label{tab:nodecontractions}
        \end{table*}
By examining the diagrammatic contractions from Table \ref{tab:nodecontractions}, and maintaining the symmetry between the pairs $(ab)(st)$ and $(cd)(uv)$, we conclude that the relevant contractions are $\Delta^2_{ab,cd,st,uv}$, $\left(\Delta^1_{ab,cd,st,uv}+\Delta^3_{ab,cd,st,uv}\right)$, $\Delta^5_{ab,cd,st,uv}$ and $\left(\Delta^4_{ab,cd,st,uv}+\Delta^6_{ab,cd,st,uv}\right)$. This is evident because switching two symmetry-related pairs [that is, switching $(ab)$ and $(st)$ or $(cd)$ and $(su)$] in diagram 1 yields diagram 3. Similarly, switching two symmetry-related pairs in diagram 4 yields diagram 6.
We see that there are four contractions but we have only three parameters.
This implies that out of these four contractions there are only three linearly independent contractions.

We can now write down a general interaction Hamiltonian $V_{ab,cd,st,uv}$ as a linear combination of three Kronecker delta products, each multiplied by one of the three independent coefficients:
\begin{align}
    \nonumber
        V_{ab,cd,st,uv}^{\phantom{2}} &= c_0 \Delta^2_{ab,cd,st,uv}\\
    \nonumber
        & + c_1 \left(\Delta^1_{ab,cd,st,uv}+\Delta^3_{ab,cd,st,uv}\right) \\
    \label{eq:Vabcdstuv_tensor}
        &+ c_2 \Delta^5_{ab,cd,st,uv}.
\end{align}
We find $V_{\alpha\beta,\chi\gamma}$ by transforming this expression back to the five-dimensional fourth-rank representation, as dictated by Eq.~(\ref{eq:V_dorb_irrep}).  For completeness we include all six of the transformed contractions:
\begin{align}
    \nonumber \sum_{abcd}\sum_{stuv}&\Delta_{ab,cd,st,uv}^{1}\xi_{\alpha ab}^{\phantom{1}}\xi_{\beta cd}^{\phantom{1}}\xi_{\chi st}^{\phantom{1}}\xi_{\gamma uv}^{\phantom{1}}=\\
    \nonumber
        &\sum_{abst}4\xi_{\alpha ab}^{\phantom{1}}\xi_{\beta ba}^{\phantom{1}}\xi_{\chi st}^{\phantom{1}}\xi_{\gamma ts}^{\phantom{1}}=\delta_{\alpha\beta}^{\phantom{1}}\delta_{\chi\gamma}^{\phantom{1}},\\
    \nonumber \sum_{abcd}\sum_{stuv}&\Delta_{ab,cd,st,uv}^{2}\xi_{\alpha ab}^{\phantom{2}}\xi_{\beta cd}^{\phantom{2}}\xi_{\chi st}^{\phantom{2}}\xi_{\gamma uv}^{\phantom{2}}=\\
    \nonumber
        &\sum_{abcd}4\xi_{\alpha ab}^{\phantom{2}}\xi_{\chi ba}^{\phantom{2}}\xi_{\beta cd}^{\phantom{2}}\xi_{\gamma dc}^{\phantom{2}}=\delta_{\alpha\chi}^{\phantom{2}}\delta_{\beta\gamma}^{\phantom{2}},\\
    \nonumber \sum_{abcd}\sum_{stuv}&\Delta_{ab,cd,st,uv}^{3}\xi_{\alpha ab}^{\phantom{3}}\xi_{\beta cd}^{\phantom{3}}\xi_{\chi st}^{\phantom{3}}\xi_{\gamma uv}^{\phantom{3}}=\\
    \nonumber
        &\sum_{abcd}4\xi_{\alpha ab}^{\phantom{3}}\xi_{\gamma ba}^{\phantom{3}}\xi_{\beta cd}^{\phantom{3}}\xi_{\chi dc}^{\phantom{3}}=\delta_{\alpha\gamma}^{\phantom{3}}\delta_{\beta\chi}^{\phantom{3}},\\
    \nonumber \sum_{abcd}\sum_{stuv}&\Delta_{ab,cd,st,uv}^{4}\xi_{\alpha ab}^{\phantom{4}}\xi_{\beta cd}^{\phantom{4}}\xi_{\chi st}^{\phantom{4}}\xi_{\gamma uv}^{\phantom{4}}=\\
    \nonumber
        &\sum_{abdv}16\xi_{\alpha ab}^{\phantom{4}}\xi_{\beta bd}^{\phantom{4}}\xi_{\gamma dv}^{\phantom{4}}\xi_{\chi va}^{\phantom{4}},\\
    \nonumber \sum_{abcd}\sum_{stuv}&\Delta_{ab,cd,st,uv}^{5}\xi_{\alpha ab}^{\phantom{5}}\xi_{\beta cd}^{\phantom{5}}\xi_{\chi st}^{\phantom{5}}\xi_{\gamma uv}^{\phantom{5}}=\\
    \nonumber
        &\sum_{abdt}16\xi_{\alpha ab}^{\phantom{5}}\xi_{\beta bd}^{\phantom{5}}\xi_{\chi dt}^{\phantom{5}}\xi_{\gamma ta}^{\phantom{5}},\\
    \nonumber
        \sum_{abcd}\sum_{stuv}&\Delta_{ab,cd,st,uv}^{6}\xi_{\alpha ab}^{\phantom{5}}\xi_{\beta cd}^{\phantom{5}}\xi_{\chi st}^{\phantom{5}}\xi_{\gamma uv}^{\phantom{5}}=\\
    \label{eq:xicontractions}
        &\sum_{abdt}16\xi_{\alpha ab}^{\phantom{5}}\xi_{\chi bt}^{\phantom{5}}\xi_{\beta td}^{\phantom{5}}\xi_{\gamma da}^{\phantom{5}}.
\end{align}
We find that the last three of the transformations of the contractions, although rotationally invariant in orbital space, are not expressible in terms of Kronecker deltas of $\alpha$, $\beta$, $\chi$ and $\gamma$. This is to be expected as there exist terms in $V_{\alpha\beta,\chi\gamma}$ that are non-zero and have more than two orbital indices, e.g.~$V_{(3z^2-r^2)(xy),(yz) (xz)}$ or $V_{(3z^2-r^2)(xy),(xy)(x^2-y^2)}$.

We can now write down $V_{\alpha\beta,\chi\gamma}$ concisely,
\begin{align}
    \nonumber
        V_{\alpha\beta,\chi\gamma}&=U\delta_{\alpha\chi}\delta_{\beta\gamma}+\left(J+\frac{5}{2}\Delta J\right)(\delta_{\alpha\gamma}\delta_{\beta\chi}+\delta_{\alpha\beta}\delta_{\chi\gamma})\\
    \label{eq:V_dorb_tensor}
        &-48\Delta J\sum_{abdt}\xi_{\alpha ab}\xi_{\beta bd}\xi_{\chi dt}\xi_{\gamma ta},
\end{align}
where we have defined $U$ to be the Hartree term between the~$t_{2g}$ orbitals on-site, $U = V_{(zx)(yz),(zx)(yz)}$, $J$ to be the average of the exchange integral between the $e_g$ and $t_{2g}$ orbitals on-site, $J=\frac{1}{2}\left(V_{(zx)(yz),(yz)(zx)}+V_{(3z^{2}-r^2)(x^{2}-y^{2}),(x^{2}-y^{2})(3z^{2}-r^2)}\right)$, and $\Delta J$ to be the difference between the exchange integrals for $e_g$ and $t_{2g}$, $\Delta J=V_{(3z^{2}-r^2)(x^{2}-y^{2}),(x^{2}-y^{2})(3z^{2}-r^2)}-V_{(zx)(yz),(yz)(zx)}$. We could equally well have chosen to use the sum of the fourth and sixth contractions instead of the fifth contraction, although with different coefficients.

\section{Angular momentum and quadrupole operators}
\label{append:quadrupole}
\subsection{Angular momentum}
\label{subappend:angmom}
The angular momentum operators are most naturally expressed in spherical polar coordinates, even when using them to operate on cubic harmonics. Their forms are:
\begin{align}
    \label{eq:Lx}
        \hat{L}_x &= i \left(\sin \phi\frac{\6}{\6 \theta} + \cot \theta \cos\phi \frac{\6}{\6 \phi}\right),\\
    \label{eq:Ly}
        \hat{L}_y &= i \left(-\cos\phi \frac{\6}{\6 \theta}+\cot\theta\sin\phi\frac{\6}{\6 \phi}\right),\\
    \label{eq:Lz}
        \hat{L}_z &= -i \frac{\6}{\6\phi},
\end{align}
where we have set $\hbar$ to 1 for simplicity.
When applying the above operators to the on-site cubic harmonic p orbitals (we have dropped the site index for clarity) it is straightforward to show that 
\begin{equation}
    \label{eq:L_p_orb}
        \hat{L}_{\mu}p_j=i \sum_{k}\epsilon_{\mu jk} p_k.
\end{equation}
A general one particle operator $\hat{O}$ may be expressed in terms of creation and annihilation operators as follows:
\begin{equation}
    \label{eq:operator_def}
        \hat{O}=\sum_{\alpha\beta\sigma\sigma'}\langle\phi^{\phantom{\dagger}}_{\alpha\sigma}|\hat{O}|\phi^{\phantom{\dagger}}_{\beta\sigma'}\rangle\hat{c}^{\dagger}_{\alpha\sigma}\hat{c}^{\phantom{\dagger}}_{\beta\sigma'}.
\end{equation}
Substituting Eq.~(\ref{eq:L_p_orb}) into Eq.~(\ref{eq:operator_def}) yields
\begin{equation}
    \label{eq:L_p_op}
        \hat{L}_{\mu}=i\sum_{\alpha\beta\sigma} \epsilon_{\mu\beta\alpha}\hat{c}^{\dagger}_{\alpha\sigma}\hat{c}^{\phantom{\dagger}}_{\beta\sigma},
\end{equation}
where $\mu$ is a Cartesian direction and $\alpha$ and $\beta$ are cubic harmonic p orbitals ($x$, $y$ or $z$). The d case is slightly more complicated, but is greatly simplified by the use of Eq.~(\ref{eq:dorb_trans_irrep_cub}). Applying the angular momentum operators to the $B$ matrix reveals that
\begin{align}
    \label{eq:L_B_mat}
        \hat{L}_{\mu } |B_{jk}\rangle &= i\sum_s\left(\epsilon_{\mu js}|B_{sk}\rangle+\epsilon_{\mu ks}|B_{js}\rangle\right),
\end{align}
with
\begin{align}
    \label{eq:L_d_orb}
        \hat{L}_{\mu } |\phi_{\alpha\sigma}\rangle & = i N_d \sum_{jks}S(\sigma)\frac{R_d(r)}{r^2} \xi_{\alpha jk}\left(\epsilon_{\mu js}|B_{sk}\rangle+\epsilon_{\mu ks}|B_{js}\rangle\right).
\end{align}
By finding the expectation value and substituting into Eq.~(\ref{eq:operator_def}) we obtain 
\begin{equation}
    \label{eq:L_d_op}
        \hat{L}_{\mu}=4i\sum_{jkm}\sum_{\alpha\beta\sigma}\epsilon^{\phantom{\dagger}}_{\mu m k} \xi^{\phantom{\dagger}}_{\alpha k j}\xi^{\phantom{\dagger}}_{\beta j m}\hat{c}^{\dagger}_{\alpha\sigma}\hat{c}^{\phantom{\dagger}}_{\beta\sigma}.
\end{equation}

\subsection{Quadrupole operator} \label{subappend:quadrupole} 

We define the quadrupole operator for a single electron as
\begin{equation}
\label{eq:quadrupole_op} \hat{Q}_{\mu
\nu}=\frac{1}{2}\left(\hat{L}_\mu\hat{L}_\nu+\hat{L}_\nu\hat{L}_\mu\right)-\frac{1}{3}\delta_{\mu
\nu}\hat{\mbf L}^2.
\end{equation}
Here we interpret $\hat{L}_{\mu}$ and $\hat{L}_{\nu}$ in Eq.~(\ref{eq:quadrupole_op}) as components of the angular momentum of a single electron.
The quadrupole operator for a system of $N$ electrons is then a sum over contributions from each electron: $\hat{Q}_{\mu\nu} = \sum_{i=1}^{N} \hat{Q}_{\mu\nu}(i)$.
This makes $\hat{Q}_{\mu\nu}$ a one-electron operator, respresented in second quantization as a linear combination of strings of one creation operator and one annihilation operator.
Squaring and tracing the one-electron tensor operator $\hat{Q}_{\mu\nu}$ yields a two-electron operator $\hat{Q}^2 = \sum_{\mu\nu} \hat{Q}_{\mu\nu} \hat{Q}_{\nu\mu}$.

To find the form of this operator, we start by applying the single electron version of $\hat{Q}_{\mu \nu}$ to the $B$ matrix, making use of Eq.~(\ref{eq:L_B_mat}).
Starting with $\hat{L}_\mu\hat{L}_\nu$ 
\begin{align} \nonumber
\hat{L}_\mu\hat{L}_\nu|B_{jk}\rangle &= i\hat{L}_\mu\sum_s\left(\epsilon_{\nu
js}|B_{sk}\rangle+\epsilon_{\nu ks}|B_{js}\rangle\right),\\ \nonumber 
&= (i)^2\sum_{ss'}\bigg(\epsilon_{\nu js}\left(\epsilon_{\mu
ss'}|B_{s'k}\rangle+\epsilon_{\mu ks'}|B_{ss'}\rangle\right)\\ \nonumber
&+\epsilon_{\nu ks}\left(\epsilon_{\mu js'}|B_{s's}\rangle+\epsilon_{\mu
ss'}|B_{js'}\rangle\right)\bigg),\\ \nonumber &=(-1)\bigg(\delta_{j\mu}|B_{\nu
k}\rangle+\delta_{k\mu}|B_{j\nu}\rangle-2\delta_{\nu \mu}|B_{jk}\rangle\\
\label{eq:L_mu_L_nu} &+\sum_{ss'}\left(\epsilon_{\nu js}\epsilon_{\mu
ks'}+\epsilon_{\nu ks}\epsilon_{\mu js'}\right)|B_{ss'}\rangle\bigg),\\
\nonumber \hat{L}_\nu\hat{L}_\mu|B_{jk}\rangle &=(-1)\bigg(\delta_{j\nu}|B_{\mu
k}\rangle+\delta_{k\nu}|B_{j\mu}\rangle-2\delta_{\mu \nu}|B_{jk}\rangle\\
\label{eq:L_nu_L_mu} &+\sum_{ss'}\left(\epsilon_{\mu js}\epsilon_{\nu
ks'}+\epsilon_{\mu ks}\epsilon_{\nu js'}\right)|B_{ss'}\rangle\bigg),
\end{align}
where the last equation was found by exchanging $\mu$ and $\nu$ in the previous equation.
Following the same process, the final term in Eq.~(\ref{eq:quadrupole_op}) becomes
\begin{equation}
\label{eq:SumpLpLpBmat} \sum_{\nu
}\hat{L}_\nu\hat{L}_\nu|B_{jk}\rangle=6|B_{jk}\rangle, \end{equation} 
This is not a surprising result as the $B$ matrix contains linear combinations of spherical harmonics $|l,l_z\rangle$ with $l=2$, and $\hat{\mbf L}^2|l,l_z\rangle=l(l+1)|l,l_z\rangle$.
Combining Eqs.~(\ref{eq:L_mu_L_nu}), (\ref{eq:L_nu_L_mu})
and (\ref{eq:SumpLpLpBmat}) we find
\begin{align}
\nonumber \hat{Q}_{\mu \nu}|B_{jk}\rangle&=-\bigg(\frac{1}{2}\left(\delta_{\mu j}|B_{\nu k}\rangle+\delta_{k\mu}|B_{j\nu}\rangle \right . \\ \notag
& \qquad\quad \left . +\delta_{j\nu}|B_{\mu
k}\rangle+\delta_{k\nu}|B_{j\mu}\rangle\right)\\ \label{eq:quadrupole_op_2}
&+\sum_{ss'}\left(\epsilon_{\nu js}\epsilon_{\mu ks'}+\epsilon_{\nu
ks}\epsilon_{\mu js'}\right)|B_{ss'}\rangle\bigg).
\end{align}
Equation~(\ref{eq:operator_def}) yields the corresponding operator for a system of many electrons:
\begin{align}
\nonumber \hat{Q}_{\mu \nu}&=-\sum_{\alpha\beta\sigma}
\bigg(2\sum_{k}\left(\xi_{\alpha \nu k}^{\phantom{\dagger}} \xi_{\beta k
\mu}^{\phantom{\dagger}}+\xi_{\alpha \mu k}^{\phantom{\dagger}}\xi_{\beta
k\nu}^{\phantom{\dagger}}\right)\\ \label{eq:quadrupole_op_3}
&+4\sum_{mnjk}\xi_{\alpha mn}^{\phantom{\dagger}}\xi_{\beta
jk}^{\phantom{\dagger}}\epsilon_{\nu jm}^{\phantom{\dagger}} \epsilon_{\mu
kn}^{\phantom{\dagger}}\bigg)\hat{c}^{\dagger}_{\alpha\sigma}\hat{c}^{\phantom{\dagger}}_{\beta\sigma}.
\end{align}
We now define the normal ordered quadrupole squared as
\begin{equation} \label{eq:quadrupole_sq_op}
:\hat{Q}^2:=\sum_{\mu\nu}:\hat{Q}_{\mu \nu}\hat{Q}_{\nu \mu}:.
\end{equation}
By substituting Eq.~(\ref{eq:quadrupole_op_3}) into Eq.~(\ref{eq:quadrupole_sq_op}) and dropping the site index for clarity, one
obtains the following simple formula for the quadrupole squared operator:
\begin{align} 
\nonumber :\hat{Q}^2:&=\sum_{\alpha\beta\gamma\chi\sigma\sigma'}\bigg(-3
\delta_{\alpha\chi}^{\phantom{\dagger}}\delta_{\beta\gamma}^{\phantom{\dagger}}+9\left(\delta_{\alpha\gamma}^{\phantom{\dagger}}\delta_{\beta\chi}^{\phantom{\dagger}}+\delta_{\alpha\beta}^{\phantom{\dagger}}\delta_{\chi\gamma}^{\phantom{\dagger}}\right)\\
\label{eq:quadrupole_sq_op_2} &-72\sum_{stuv}\xi_{\alpha s
t}^{\phantom{\dagger}}\xi_{\beta t u}^{\phantom{\dagger}}\xi_{\chi u
v}^{\phantom{\dagger}}\xi_{\gamma v
s}^{\phantom{\dagger}}\bigg)\hat{c}^{\dagger}_{\alpha\sigma}\hat{c}^{\dagger}_{\beta\sigma'}\hat{c}^{\phantom{\dagger}}_{\gamma\sigma'}\hat{c}^{\phantom{\dagger}}_{\chi\sigma}.
\end{align}

\subsection{The quadrupole operator squared and the on-site Coulomb integrals}
\label{subappend:quadrupole_onsite}
The quadrupole operator is of interest to us because one of the terms in the d-shell on-site Coulomb interaction can be represented in terms of $\hat{Q}^2$. This term is:
\begin{equation}
-48\Delta J\sum_{\alpha\beta\chi\gamma\sigma\sigma'}\sum_{stuv}\xi_{\alpha s t}^{\phantom{\dagger}}\xi_{\beta t u}^{\phantom{\dagger}}\xi_{\chi u v}^{\phantom{\dagger}}\xi_{\gamma v s}^{\phantom{\dagger}}\hat{c}^{\dagger}_{\alpha\sigma}\hat{c}^{\dagger}_{\beta\sigma'}\hat{c}^{\phantom{\dagger}}_{\gamma\sigma'}\hat{c}^{\phantom{\dagger}}_{\chi\sigma},
\end{equation}
which is proportional to the final term in Eq.~(\ref{eq:quadrupole_sq_op_2}). The other terms in Eq.~(\ref{eq:quadrupole_sq_op_2}) also exist elsewhere in the d-shell on-site interaction Hamiltonian, so introducing an explicit $\hat{Q}^2$ term is straightforward:
\begin{align}
    \nonumber
        -&48 \Delta J \sum_{\alpha\beta\chi\gamma\sigma\sigma'}\sum_{stuv}\xi_{\alpha st}^{\phantom{\dagger}}\xi_{\beta tu}^{\phantom{\dagger}} \xi_{\chi uv}^{\phantom{\dagger}}\xi_{\gamma vs}^{\phantom{\dagger}}\hat{c}^\dagger_{\alpha\sigma}\hat{c}^{\dagger}_{\beta\sigma'}\hat{c}^{\phantom{\dagger}}_{\gamma\sigma'}\hat{c}^{\phantom{\dagger}}_{\chi\sigma} \\
    \nonumber
        &= \frac{2}{3}\Delta J:\hat{Q}^2:+2\Delta J \sum_{\alpha\beta\sigma\sigma'}\hat{c}^\dagger_{\alpha\sigma}\hat{c}^{\dagger}_{\beta\sigma'}\hat{c}^{\phantom{\dagger}}_{\beta\sigma'}\hat{c}^{\phantom{\dagger}}_{\alpha\sigma}\\
    \nonumber
        &-6\Delta J \sum_{\alpha\beta\sigma\sigma'}(\hat{c}^\dagger_{\alpha\sigma}\hat{c}^{\dagger}_{\beta\sigma'}\hat{c}^{\phantom{\dagger}}_{\alpha\sigma'}\hat{c}^{\phantom{\dagger}}_{\beta\sigma}+\hat{c}^\dagger_{\alpha\sigma}\hat{c}^{\dagger}_{\alpha\sigma'}\hat{c}^{\phantom{\dagger}}_{\beta\sigma'}\hat{c}^{\phantom{\dagger}}_{\beta\sigma})\\
    \nonumber
        &= \frac{2}{3}\Delta J:\hat{Q}^2:+2\Delta J:\hat{n}^{2}:+3\Delta J(:\hat{n}^{2}:+:\hat{{\mbf m}}^{2}:)\\
    \label{eq:insert_quadrupole}
        &-6\Delta J\sum_{\alpha\beta}:\left(\hat{n}^{\phantom{\dagger}}_{\alpha\beta}\right)^2:.
\end{align}

\bibliographystyle{jabbrv_apalike}
\bibliography{mybiblio}

\end{document}